\documentclass[reprint,superscriptaddress,amssymb,amsmath,aps,longbibliography,nofootinbib, pra]{revtex4-2}
\usepackage{graphicx}
\usepackage{dcolumn}
\usepackage{bm}
\usepackage{float}
\usepackage{comment}
\usepackage{amsmath}
\usepackage{amssymb}
\usepackage{color}
\usepackage{subeqnarray}
\usepackage[utf8]{inputenc}
\usepackage[english]{babel}
\usepackage{mathrsfs}
\usepackage{float}
\usepackage[bb=boondox]{mathalfa}

\usepackage{amsbsy}
\usepackage{latexsym,epsfig,graphicx}
\usepackage{subfigure}
\usepackage{amsfonts}
\usepackage{amsmath}
\usepackage{xspace}
\usepackage{epstopdf}
\usepackage{array,booktabs}
\usepackage{babel}
\usepackage{multirow}
\usepackage[colorlinks=true, letterpaper=true, pdfstartview=FitV, linkcolor=red, citecolor=blue, urlcolor=blue]{hyperref}
\usepackage[normalem]{ulem}
\usepackage{cleveref,braket,xcolor}

\pdfoutput=1

\begin{document}

\title{Quantum Langevin theory for two coupled phase-conjugated electromagnetic waves}

\author{Yue Jiang}
\email{yue.jiang@jila.colorado.edu}
\affiliation{JILA, National Institute of Standards and Technology and the University of Colorado, Boulder, Colorado 80309, USA}
\affiliation{Department of Physics, University of Colorado, Boulder, Colorado 80309, USA}

\author{Yefeng Mei}
\email{meiyf@umich.edu}
\affiliation{Department of Physics, University of Michigan, Ann Arbor, Michigan 48109, USA}

\author{Shengwang Du}
\email{dusw@utdallas.edu}
\affiliation{Department of Physics, The University of Texas at Dallas, Richardson, Texas 75080, USA}

\date{\today}

\begin{abstract}
While loss-gain-induced Langevin noises have been intensively studied in quantum optics, the effect of a complex-valued nonlinear coupling coefficient on the noises of two coupled phase-conjugated optical fields has never been questioned before. Here, we provide a general macroscopic phenomenological formula of quantum Langevin equations for two coupled phase-conjugated fields with linear loss (gain) and complex nonlinear coupling coefficient. The macroscopic phenomenological formula is obtained from the coupling matrix to preserve the field commutation relations and correlations, which does not require knowing the microscopic details of light-matter interaction and internal atomic structures. To validate this phenomenological formula, we take spontaneous four-wave mixing in a double-$\Lambda$ four-level atomic system as an example to numerically confirm that our macroscopic phenomenological result is consistent with that obtained from the microscopic Heisenberg-Langevin theory. Finally, we apply the quantum Langevin equations to study the effects of linear gain and loss, complex phase mismatching, as well as complex nonlinear coupling coefficient in entangled photon pair (biphoton) generation, particularly to their temporal quantum correlations.
\end{abstract}

\maketitle


\section{Introduction}\label{sec:Introduction}

Quantum Langevin equations is a common approach to studying an open quantum system involving loss or gain, where the stochastic coupling between the system and its environment is molded as a set of Langevin noise operators \cite{gardiner1985input, scully1999quantum, yamamoto1999mesoscopic, gardiner2004quantum, PhysRevLett.46.1}. For example, in the parametric down-conversion (PDC) process, a pump laser beam passes through a $\chi^{(2)}$ nonlinear crystal and is down-converted into a pair of phase-conjugated electromagnetic (EM) waves. In the simplest case with the perfect phase-matching condition and an undepleted pump beam, without linear loss or gain, the two phase-conjugated single-mode fields are governed by the following coupled equations \cite{boyd2020nonlinear}
\begin{equation}
\begin{aligned}
\frac{\partial}{\partial z}\left[\begin{matrix}\hat a_{1}\\\hat a_2^\dag\\\end{matrix}\right]
=\mathrm{M} \left[\begin{matrix}\hat a_{1}\\{\hat a_2^{\dag}}\\\end{matrix}\right]
=\left[\begin{matrix}0&i\kappa\\-i\kappa&0\\\end{matrix}\right]\left[\begin{matrix}\hat a_{1}\\{\hat a_2^{\dag}}\\\end{matrix}\right],
\end{aligned}\label{eq:PDC0}
\end{equation}
where $\hat a_{m}$ and $\hat a_{m}^{\dag}$ ($m=1,2$) are the field annihilation and creation operators, $\mathrm{M}$ is the $2\times 2$ coupling matrix, and $\kappa$ is the (real) nonlinear coupling coefficient. Here we consider only the forward-wave case with both fields propagating along the same $+z$ direction. If losses are presented during the propagation of the two fields, the coupling matrix is 
\begin{equation}
\begin{aligned}
\mathrm{M}=\left[\begin{matrix}-\alpha_1&i\kappa\\-i\kappa&-\alpha_2\\\end{matrix}\right],
\end{aligned}\label{eq:M1}
\end{equation}
and their coupled equations become \cite{shwartz2012x,yamamoto1999mesoscopic}
\begin{equation}
\begin{aligned}
\frac{\partial}{\partial z}\left[\begin{matrix}\hat a_{1}\\\hat a_2^\dag\\\end{matrix}\right]
=\left[\begin{matrix}-\alpha_1&i\kappa\\-i\kappa&-\alpha_2\\\end{matrix}\right]\left[\begin{matrix}\hat a_{1}\\{\hat a_2^{\dag}}\\\end{matrix}\right]+\left[\begin{matrix}\sqrt{2\alpha_1}\hat f_{1}\\{\sqrt{2\alpha_2}\hat f_2^{\dag}}\\\end{matrix}\right],
\end{aligned}\label{eq:PDC1}
\end{equation}
where $\alpha_m > 0$ are the loss (absorption) coefficients, and $\hat f_m$ are the associated Langevin noise operators satisfying $[\hat f_m(\omega, z),\hat f_n^{\dag}(\omega', z')]=\delta_{mn}\delta(\omega-\omega')\delta(z-z')$. If there is linear gain instead of loss, for example in channel 1, \textit{i.e.}, $\alpha_1<0$, equation \eqref{eq:PDC1} can be modified by taking $\sqrt{2\alpha_1}\hat f_{1} \rightarrow \sqrt{-2\alpha_1}\hat f_{1}^{\dag}$. One can show that these Langevin noise operators are necessary to preserve the commutation relations during propagation, \textit{i.e.} $[\hat a_m(\omega, z),\hat a_n^{\dag}(\omega', z)]=[\hat a_m(\omega, 0),\hat a_n^{\dag}(\omega', 0)]=\delta_{mn}\delta(\omega-\omega')$.

Equation \eqref{eq:PDC1} has been widely applied for PDC processes where the nonlinear coupling coefficient $\kappa$ is real \cite{shwartz2012x,yamamoto1999mesoscopic, PhysRevA.100.043811, Shafiee_2020}. However, in a more general case of coupled phase-conjugated fields, such as four-wave mixing (FWM) near atomic resonances \cite{du2008narrowband, kolchin2007electromagnetically, zhao2016narrowband}, the nonlinear coupling coefficient $\kappa$ can take a complex value involving complicated atomic transitions. In this case, equation \eqref{eq:PDC1} is not valid and its solution does not preserve commutation relations of the fields. What are the general quantum Langevin coupled equations accounting for the complex nonlinear coupling coefficient? 

To answer the question, the common approach is to derive quantum Langevin equations by solving the light-matter coupled Heisenberg equations, which requires knowing microscopic details of light-matter interaction such as atomic populations and transitions \cite{kolchin2007electromagnetically, ooi2007correlation, zhao2016narrowband}. The complexity of this approach increases dramatically as more atomic transitions are involved and it is extremely difficult for experimentalists to follow, particularly in some situations where it is impossible to obtain full microscopic details. Then our reduced question becomes: Is it possible to obtain self-consistent quantum Langevin coupled equations from the general expression of the coupling matrix? We call this the macroscopic phenomenological approach. To our best knowledge, there has been no published work in investigating Langevin noises induced by a complex nonlinear coupling coefficient $\kappa$. 

In this article, for the first time, we provide a general macroscopic phenomenological formula of quantum Langevin equations for two coupled phase-conjugated fields with linear loss (gain) and complex nonlinear coupling coefficient, in both forward- and backward-wave configurations. The macroscopic phenomenological formula is obtained from the coupling matrix by preserving commutation relations and correlations of the fields, which does not require knowing the microscopic details of light-matter interaction and internal atomic structures. We aim to make it readable and accessible for experimental researchers in the quantum optics community. 

This article is structured as follows. In Sec.~\ref{sec:QLE}, to fulfill the requirement of preserving commutation relations, we formulate the general macroscopic phenomenological quantum Langevin coupled equations and their solutions from the coupling matrix taking into account linear loss (gain) and complex nonlinear coupling coefficient, in both forward- and backward-wave configurations. In Sec.~\ref{sec:Microscopic}, taking spontaneous four-wave mixing (SFWM) in a double-$\Lambda$ four-level atomic system as an example, we derive the coupled Langevin equations from microscopic light-atom Heisenberg interaction for this special case. We numerically confirm that the macroscopic phenomenological solution in Sec.~\ref{sec:QLE} agrees well with the microscopic approach. In Sec.~\ref{sec:Biphoton}, we apply the quantum Langevin theory to study effects of linear gain and loss, complex phase mismatching, and complex nonlinear coupling coefficient in entangled photon pair (biphoton) generation, particularly to their temporal quantum correlations. We conclude in the last section \ref{sec:Conclusion}.

\begin{figure*}[t]
\centering
\includegraphics[width=0.8 \textwidth]{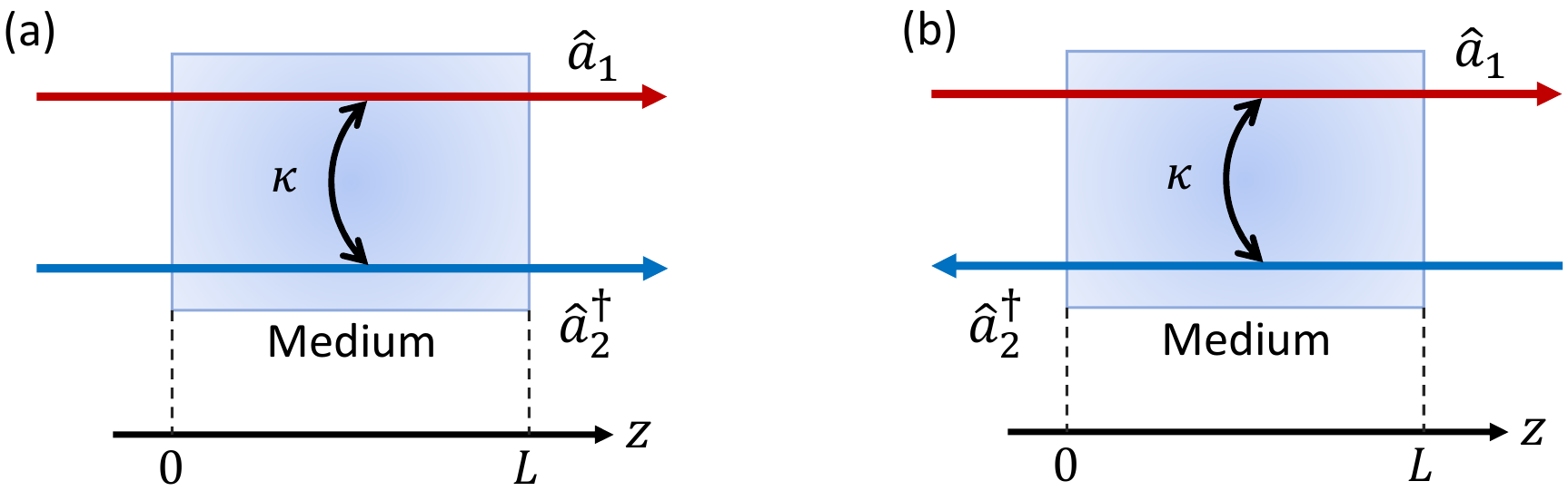}
\caption{Schematics of two coupled phase-conjugated electromagnetic waves: (a) forward-wave configuration, and (b) backward-wave configuration. $\kappa$ is the nonlinear coupling coefficient between the two modes.}
\label{fig:Fig01}
\end{figure*}

\section{Quantum Langevin Equations}\label{sec:QLE}

Here we consider the two coupled single-mode phase-conjugated fields in either forward-wave or backward-wave configuration, as illustrated in Fig.~\ref{fig:Fig01}. In the forward-wave configuration [Fig.~\ref{fig:Fig01}(a)], both fields propagate along $+z$ direction through a nonlinear medium with a length $L$. In the backward-wave configuration [Fig.~\ref{fig:Fig01}(b)], the two fields propagate in opposing directions. The field annihilation operators $\hat a_m(t,z)$ can be expressed as 
\begin{equation}
\begin{aligned}
&\hat a_1(t,z)
=\frac{1}{\sqrt{2\pi}}\int{d\omega \hat a_1(\omega,z)e^{i(\frac{\omega}{c}z-\omega t)}},\\
&\hat a_2(t,z)
=\frac{1}{\sqrt{2\pi}}\int{d\omega \hat a_2(\omega,z)e^{i(\pm \frac{\omega}{c}z-\omega t)}},
\end{aligned}
\label{eq:Fields}
\end{equation}
where $\pm$ represents that field 2 propagates along $+z$ or $-z$ direction, for the forward-wave or backward-wave configuration, respectively. The filed operators satisfy the following commutation relations
\begin{equation}
\begin{aligned}
&\left[\hat a_m\left(t,z\right),\hat a_n^\dag\left(t^\prime,z\right)\right]=\delta_{mn}\delta(t-t^\prime),\\
&\left[\hat a_m\left(\omega,z\right),\hat a_n^\dag\left(\omega^\prime,z\right)\right]=\delta_{mn}\delta(\omega-\omega^\prime).
\end{aligned}
\label{eq:commutation0}
\end{equation}

In the forward-wave configuration, both fields are input at $z=0$, or $\hat a_1(0)$ and $\hat a_2(0)$ are the ``initial'' boundary conditions. The general coupling matrix is \cite{PhysRevLett.123.193604}
\begin{equation}
\begin{aligned}
\mathrm{M_{F}}=\left[\begin{matrix}-\alpha_1+i\frac{\Delta k}{2}&i\kappa\\-i\kappa&-\alpha_2^*-i\frac{\Delta k}{2}\\\end{matrix}\right],
\end{aligned}\label{eq:MFW}
\end{equation}
where $\alpha_m=-i\frac{\omega_m}{2c}\chi_m$ with $\chi_m$ being linear susceptibility, and $\Delta k$ (real) is the phase mismatching in vacuum. In general, $\alpha_m$ is complex valued, whose real part $\mathrm{Re}\{\alpha_m\}>0$ represents loss (or gain for $\mathrm{Re}\{\alpha_m\}<0$) and imaginary part represents phase velocity dispersion. The nonlinear coupling coefficient $\kappa$ can also be complex-valued. In the backward-wave configuration, the general coupling matrix becomes \cite{zhao2016narrowband,mei2017mirrorless}
\begin{equation}
\begin{aligned}
\mathrm{M_{B}}=\left[\begin{matrix}-\alpha_1+i\frac{\Delta k}{2}&i\kappa\\i\kappa&\alpha_2^*-i\frac{\Delta k}{2}\\\end{matrix}\right],
\end{aligned}\label{eq:MBW}
\end{equation}
and the ``initial'' boundary conditions are $\hat a_1(0)$ and $\hat a_2(L)$: field 1 is input at $z=0$ and field 2 is input at $z=L$.   

One can show that, under the following unitary gauge transformation
\begin{equation}
\begin{aligned}
\left[\begin{matrix}\hat a_{1}\\\hat a_2^\dag\\\end{matrix}\right]
=\left[\begin{matrix}e^{i\theta/2}&0\\0&e^{-i\theta/2}\\\end{matrix}\right]\left[\begin{matrix}\hat a_{1}\\\hat a_2^\dag\\\end{matrix}\right]
=\mathrm{U}\left[\begin{matrix}\hat a_{1}\\\hat a_2^\dag\\\end{matrix}\right]=\left[\begin{matrix}\hat a_{1}e^{i\theta/2}\\\hat a_2^\dag e^{-i\theta/2}\\\end{matrix}\right],
\end{aligned}\label{eq:GT1}
\end{equation}
the corresponding coupling matrix become
\begin{equation}
\begin{aligned}
\mathrm{M_{F}(\theta)}=\mathrm{U}\mathrm{M_{F}}\mathrm{U}^{\dag}=\left[\begin{matrix}-\alpha_1+i\frac{\Delta k}{2}&i\kappa e^{i\theta}\\-i\kappa e^{-i\theta}&-\alpha_2^*-i\frac{\Delta k}{2}\\\end{matrix}\right],
\end{aligned}\label{eq:MFWU}
\end{equation}
and
\begin{equation}
\begin{aligned}
\mathrm{M_{B}(\theta)}=\mathrm{U}\mathrm{M_{B}}\mathrm{U}^{\dag}=\left[\begin{matrix}-\alpha_1+i\frac{\Delta k}{2}&i\kappa e^{i\theta}\\i\kappa e^{-i\theta}&\alpha_2^*-i\frac{\Delta k}{2}\\\end{matrix}\right].
\end{aligned}\label{eq:MBWU}
\end{equation}
As physics is preserved and unchanged under the above gauge transformation, we take $\theta=0$ throughout this article for convenience and simplification.    

In presence of linear loss or gain, \textit{i.e.,} $\rm{Re}\{\alpha_m\}\neq 0$, or complex nonlinear coupling coefficient, $\kappa\neq \kappa^*$, the two-mode coupled equations must include Langevin noise operators to preserve the commutation relations of the field operators in Eq. \eqref{eq:commutation0}. The noise operators should only be related to $\rm{Re}\{\alpha_m\}$ and $\rm{Im}\{\kappa\}$. As $\kappa$ is real, the coupled equations in forward-wave configuration should be reduced to the known Eq.~\eqref{eq:PDC1}. For both forward- and backward-wave configurations in the same nonlinear material, the noise origin is the same except field 2 propagates along $\pm z$ direction for different configurations. With these guidelines, we provide quantum Langevin equations for the two phase-conjugated fields from their coupling matrix in the following subsections.  

\subsection{Forward-Wave Configuration}\label{FW system}

In the forward-wave configuration as shown in Fig.~\ref{fig:Fig01}(a), we find that its quantum Langevin coupled equations can be expressed in the following general form
\begin{equation}
\begin{aligned}
\frac{\partial}{\partial z}\left[\begin{matrix}\hat a_{1}\\\hat a_2^\dag\\\end{matrix}\right]
=\mathrm{M_{F}}\left[\begin{matrix}\hat a_{1}\\{\hat a_2^{\dag}}\\\end{matrix}\right]
+\mathrm{N_{FR}}\left[\begin{matrix}\hat f_{1}\\{\hat f_2^{\dag}}\\\end{matrix}\right]
+\mathrm{N_{FI}}\left[\begin{matrix}\hat f_{1}^\dag\\{\hat f_2}\\\end{matrix}\right]
\end{aligned}\label{eq:QLEFW}
\end{equation}
with the ``initial'' condition at $z=0$:
\begin{equation}
\begin{aligned}
\left[\hat a_m(\omega, 0), \hat a_n^{\dag} (\omega', 0)\right]=\delta_{mn}\delta(\omega-\omega').
\end{aligned}
\label{eq:InitialConditionForward}
\end{equation}
The Langevin noise operators satisfy 
\begin{equation}
\begin{aligned}
\left[\hat f_{m}(\omega, z),\hat f_{n}^\dag(\omega^\prime, z^\prime)\right]=\delta_{mn}\delta(\omega-\omega')\delta(z-z^\prime)
\end{aligned}
\label{eq:LangevinCommutation}
\end{equation}
and have the following correlations
\begin{equation}
\begin{aligned}
&\left<\hat f_{m}^\dag(\omega, z)\hat f_{n}(\omega^\prime, z^\prime)\right>=0,\\
&\left<\hat f_{m}(\omega, z)\hat f_{n}^\dag(\omega^\prime, z^\prime)\right>=\delta_{mn}\delta(\omega-\omega')\delta(z-z^\prime),\\
&\left<\hat f_{m}(\omega, z)\hat f_{n}(\omega^\prime, z^\prime)\right>=\left<\hat f_{m}^\dag(\omega, z)\hat f_{n}^\dag(\omega^\prime, z^\prime)\right>=0.
\end{aligned}\label{eq:LangevinCorrelations}
\end{equation}
The Langevin noise matrix is given by 
\begin{equation}
\begin{aligned}
\mathrm{N_F}\equiv\sqrt{-(\mathrm{M_{F}}+\mathrm{M_{F}}^\ast)}=\mathrm{N_{FR}}+i \mathrm{N_{FI}},
\end{aligned}
\label{eq:NoiseMatrixForward}
\end{equation}
where $\mathrm{N_{FR}}$ and $\mathrm{N_{FI}}$ are the real and imaginary parts of the matrix $\mathrm{N_F}$ (\textit{i.e.}, $\mathrm{N_{F}}_{mn}=\mathrm{N_{FR}}_{mn}+i\mathrm{N_{FI}}_{mn}$), respectively. 

We obtain the solution of Eq. (\ref{eq:QLEFW}) at the output surface $z=L$ as the following
\begin{equation}
\begin{aligned}
&\left[\begin{matrix}\hat a_{1}\left(L\right)\\\hat a_2^\dag\left(L\right)\\\end{matrix}\right]
=e^{\mathrm{M_{F}}L}\left[\begin{matrix}\hat a_{1}\left(0\right)\\\hat a_2^\dag\left(0\right)\\\end{matrix}\right]\\
&+\int_{0}^{L}e^{\mathrm{M_{F}}(L-z)}\left(\mathrm{N_{FR}}\left[\begin{matrix}\hat f_{1}\left(z\right)\\\hat f_2^\dag\left(z\right)\\\end{matrix}\right]+\mathrm{N_{FI}}\left[\begin{matrix}\hat f_{1}^\dag\left(z\right)\\\hat f_2\left(z\right)\\\end{matrix}\right]\right)dz.
\end{aligned}\label{eq:SolutionFW1}
\end{equation}
Defining
\begin{equation}
\begin{aligned}
e^{\mathrm{M_{F}}L}\equiv\left[\begin{matrix}A&B\\C&D\\\end{matrix}\right],
\end{aligned}\label{ABCDFW}
\end{equation}
\begin{equation}
\begin{aligned}
e^{\mathrm{M_{F}}(L-z)}\equiv\left[\begin{matrix}A_1\left(z\right)&B_1\left(z\right)\\C_1\left(z\right)&D_1\left(z\right)\\\end{matrix}\right],
\end{aligned}\label{A1B1C1D1FW}
\end{equation}
we rewrite Eq. (\ref{eq:SolutionFW1}) as
\begin{equation}
\begin{aligned}
&\left[\begin{matrix}\hat a_{1}\left(L\right)\\\hat a_2^\dag\left(L\right)\\\end{matrix}\right]
=\left[\begin{matrix}A&B\\C&D\\\end{matrix}\right]\left[\begin{matrix}\hat a_{1}\left(0\right)\\\hat a_2^\dag\left(0\right)\\\end{matrix}\right]\\
&+\int_{0}^{L}\left[\begin{matrix}A_1\left(z\right)&B_1\left(z\right)\\C_1\left(z\right)&D_1\left(z\right)\\\end{matrix}\right]\left(\mathrm{N_{FR}}\left[\begin{matrix}\hat f_{1}\left(z\right)\\\hat f_2^\dag\left(z\right)\\\end{matrix}\right]+\mathrm{N_{FI}}\left[\begin{matrix}\hat f_{1}^\dag\left(z\right)\\\hat f_2\left(z\right)\\\end{matrix}\right]\right)dz.
\end{aligned}\label{eq:SolutionFW2}
\end{equation}
We numerically confirm that the solution preserves the commutation relations 
\begin{eqnarray}
   \left[\hat a_m(\omega, L), \hat a_n^{\dag} (\omega', L)\right]&=&\left[\hat a_m(\omega, 0), \hat a_n^{\dag} (\omega', 0)\right]\nonumber\\
   &=&\delta_{mn}\delta(\omega-\omega').
\end{eqnarray}

Now we examine some special cases.

\noindent\textbf{Case 1:} We first consider the coupling matrix $\mathrm{M_{F}}$ in Eq. \eqref{eq:MFW} where the nonlinear coupling coefficient $\kappa$ is real and both modes have losses ($\mathrm{Re}\{\alpha_m\}\geq 0$) . This works for most PDC processes \cite{shwartz2012x,yamamoto1999mesoscopic}. Under such a condition, we have the following diagonalized noise matrix
\begin{equation}
\begin{aligned}
\mathrm{N_{F}}=\mathrm{N_{FR}}=\left[\begin{matrix}\sqrt{2\mathrm{Re}\{\alpha_1\}}&0\\0&\sqrt{2\mathrm{Re}\{\alpha_2\}}\\\end{matrix}\right],
\end{aligned}\label{eq:NFCase1FW}
\end{equation}
and the coupled Langevin equations
\begin{equation}
\begin{aligned}
\frac{\partial}{\partial z}\left[\begin{matrix}\hat a_{1}\\\hat a_2^\dag\\\end{matrix}\right]
=\mathrm{M_{F}}\left[\begin{matrix}\hat a_{1}\\{\hat a_2^{\dag}}\\\end{matrix}\right]
+\left[\begin{matrix}\sqrt{2\mathrm{Re}\{\alpha_1\}}\hat f_{1}\\{\sqrt{2\mathrm{Re}\{\alpha_2\}}\hat f_2^{\dag}}\\\end{matrix}\right],
\end{aligned}\label{eq:QLECase1FW}
\end{equation}
which is the well-known result in literature \cite{shwartz2012x,yamamoto1999mesoscopic}.

\noindent\textbf{Case 2:} $\kappa$ is real, the mode 1 has linear loss ($\mathrm{Re}\{\alpha_1\}=\alpha\geq 0$), and the mode 2 has linear gain ($\mathrm{Re}\{\alpha_2\}=-g\leq 0$). The noise matrix becomes
\begin{equation}
\begin{aligned}
\mathrm{N_{F}}=\left[\begin{matrix}\sqrt{2\alpha}&0\\0&i\sqrt{2 g}\\\end{matrix}\right].
\end{aligned}\label{eq:NFCase2FW}
\end{equation}
We have the following coupled Langevin equations
\begin{equation}
\begin{aligned}
\frac{\partial}{\partial z}\left[\begin{matrix}\hat a_{1}\\\hat a_2^\dag\\\end{matrix}\right]
=\mathrm{M_{F}}\left[\begin{matrix}\hat a_{1}\\{\hat a_2^{\dag}}\\\end{matrix}\right]
+\left[\begin{matrix}\sqrt{2\alpha}\hat f_{1}\\\sqrt{2 g}\hat f_2\\\end{matrix}\right].
\end{aligned}\label{eq:QLECase2FW}
\end{equation}

\noindent\textbf{Case 3:} The two modes are perfectly phase-matched without linear gain or loss: $\Delta k=0$, $\alpha_1=\alpha_2=0$, but the nonlinear coupling coefficient is complex-valued $\kappa=\eta+i\zeta$. In this case, the coupled matrix is
\begin{equation}
\begin{aligned}
\mathrm{M_{F}}=\left[\begin{matrix}0&-\zeta+i\eta\\\zeta-i\eta&0\\\end{matrix}\right].
\end{aligned}\label{eq:MFCase3}
\end{equation}
The noise matrix becomes 
\begin{equation}
\begin{aligned}
\mathrm{N_{F}}={\rm{\Theta}}(\zeta)\sqrt{\zeta}\left[\begin{matrix}1&1\\-1&1\\\end{matrix}\right]
+i{\rm{\Theta}}(-\zeta)\sqrt{-\zeta}\left[\begin{matrix}1&1\\-1&1\\\end{matrix}\right],
\end{aligned}
\label{eq:NFCase4}
\end{equation}
where ${\rm{\Theta}}(\zeta)$ is Heaviside step function, ${\rm{\Theta}}(\zeta)=1$ if $\zeta>0$, ${\rm{\Theta}}(\zeta)=0$ if $\zeta\leq0$. The Langevin coupled equations are
\begin{equation}
\begin{aligned}
\frac{\partial}{\partial z}\left[\begin{matrix}\hat a_{1}\\\hat a_2^\dag\\\end{matrix}\right]
=&\mathrm{M_{F}}\left[\begin{matrix}\hat a_{1}\\{\hat a_2^{\dag}}\\\end{matrix}\right]
+{\rm{\Theta}}(\zeta)\sqrt{\zeta}\left[\begin{matrix}1&1\\-1&1\\\end{matrix}\right]\left[\begin{matrix}\hat f_{1}\\{\hat f_2^{\dag}}\\\end{matrix}\right]\\
&+{\rm{\Theta}}(-\zeta)\sqrt{-\zeta}\left[\begin{matrix}1&1\\-1&1\\\end{matrix}\right]\left[\begin{matrix}\hat f_{1}^{\dag}\\{\hat f_2}\\\end{matrix}\right].\\
\end{aligned}\label{eq:QLECase3FW}
\end{equation}
Eq. (\ref{eq:QLECase3FW}) shows that a complex-valued nonlinear coupling coefficient also leads to Langevin noises even when there is no linear gain or loss. This is revealed by this article for the first time.

\noindent\textbf{Case 4:} As $\kappa$ is real and there is no linear loss or gain ($\alpha_1=\alpha_2=0$), the coupled equations can be written as
\begin{equation}
\begin{aligned}
i\frac{\partial}{\partial z}\left[\begin{matrix}\hat a_{1}\\\hat a_2^\dag\\\end{matrix}\right]
=\left[\begin{matrix}-\frac{\Delta k}{2}&-\kappa\\\kappa&\frac{\Delta k}{2}\\\end{matrix}\right]\left[\begin{matrix}\hat a_{1}\\{\hat a_2^{\dag}}\\\end{matrix}\right]=\hat{\mathcal{H}}\left[\begin{matrix}\hat a_{1}\\{\hat a_2^{\dag}}\\\end{matrix}\right].
\end{aligned}\label{eq:APTFWM}
\end{equation}
The effective Hamiltonian $\hat{\mathcal{H}}$ has anti-parity-time (APT) symmetry, which has been demonstrated in FWM in cold atoms \cite{PhysRevLett.123.193604, PhysRevLett.128.173602}. 

\subsection{Backward-Wave Configuration}\label{BW system}

In the back-wave configuration as shown in Fig.~\ref{fig:Fig01}(b), the quantum Langevin coupled equations can be expressed in the following general form
\begin{equation}
\begin{aligned}
\frac{\partial}{\partial z}\left[\begin{matrix}\hat a_{1}\\\hat a_2^\dag\\\end{matrix}\right]
=\mathrm{M_{B}}\left[\begin{matrix}\hat a_{1}\\{\hat a_2^{\dag}}\\\end{matrix}\right]
+\mathrm{N_{BR}}\left[\begin{matrix}\hat f_{1}\\{\hat f_2^{\dag}}\\\end{matrix}\right]
+\mathrm{N_{BI}}\left[\begin{matrix}\hat f_{1}^\dag\\{\hat f_2}\\\end{matrix}\right].
\end{aligned}\label{eq:QLEBW}
\end{equation}
Different from the forward-wave configuration, the ``boundary'' condition is
\begin{equation}
\begin{aligned}
\left[\hat a_1(\omega, 0), \hat a_1^{\dag} (\omega', 0)\right]=\left[\hat a_2(\omega, L), \hat a_2^{\dag} (\omega', L)\right]=\delta(\omega-\omega').
\end{aligned}\label{eq:InitialConditionBackward}
\end{equation}
The Langevin noise operators satisfy the same commutation relations and correlations in Eqs. \eqref{eq:LangevinCommutation} and \eqref{eq:LangevinCorrelations}. The Langevin noise matrix is given by 
\begin{equation}
\begin{aligned}
\mathrm{N_B}&\equiv\left[\begin{matrix}1&0\\0&-1\\\end{matrix}\right]\sqrt{\left[\begin{matrix}-\mathrm{M_{B11}}&-\mathrm{M_{B12}}\\\mathrm{M_{B21}}&\mathrm{M_{B22}}\\\end{matrix}\right]+\left[\begin{matrix}-\mathrm{M_{B11}}&-\mathrm{M_{B12}}\\\mathrm{M_{B21}}&\mathrm{M_{B22}}\\\end{matrix}\right]^*}\\
&=\mathrm{N_{BR}}+i \mathrm{N_{BI}},
\end{aligned}\label{eq:NoiseMatrixBackward}
\end{equation}
where $\mathrm{N_{BR}}$ and $\mathrm{N_{BI}}$ are the real and imaginary parts of the matrix $\mathrm{N_B}$, respectively. One can show that the noise matrix defined in Eq.~\eqref{eq:NoiseMatrixBackward} has the same origin as that in the forward-wave configuration in the same nonlinear material:
\begin{equation}
    \mathrm{N_B}=\left[\begin{matrix}1&0\\0&-1\\\end{matrix}\right]\mathrm{N_F}.
\end{equation}
We note that the choice of noise matrix is not unique. For example, transformation $\hat f_1\rightarrow-\hat f_1$ or/and $\hat f_2\rightarrow-\hat f_2$ does not affect computing any physical observable. We elaborate on this more in Appendix \ref{Appendix: BW Noise Matrix}.

We obtain the solution of Eq. (\ref{eq:QLEBW}) at $z=L$ as following
\begin{equation}
\begin{aligned}
&\left[\begin{matrix}\hat a_{1}\left(L\right)\\\hat a_2^\dag\left(L\right)\\\end{matrix}\right]
=e^{\mathrm{M_{B}}L}\left[\begin{matrix}\hat a_{1}\left(0\right)\\\hat a_2^\dag\left(0\right)\\\end{matrix}\right]\\
&+\int_{0}^{L}e^{\mathrm{M_{B}}(L-z)}\left(\mathrm{N_{BR}}\left[\begin{matrix}\hat f_{1}\left(z\right)\\\hat f_2^\dag\left(z\right)\\\end{matrix}\right]+\mathrm{N_{BI}}\left[\begin{matrix}\hat f_{1}^\dag\left(z\right)\\\hat f_2\left(z\right)\\\end{matrix}\right]\right)dz.
\end{aligned}\label{eq:SolutionBW1}
\end{equation}
We define
\begin{equation}
\begin{aligned}
e^{\mathrm{M_{B}}L}\equiv\left[\begin{matrix}\bar A&\bar B\\\bar C&\bar D\\\end{matrix}\right],
\end{aligned}\label{ABCDBW}
\end{equation}
\begin{equation}
\begin{aligned}
e^{\mathrm{M_{B}}(L-z)}\equiv\left[\begin{matrix}\bar A_1\left(z\right)&\bar B_1\left(z\right)\\\bar C_1\left(z\right)&\bar D_1\left(z\right)\\\end{matrix}\right].
\end{aligned}\label{A1B1C1D1BW}
\end{equation}
Different from the forward-wave case, in the backward-wave configuration, the mode 1 input is at $z=0$ and the mode 2 input is at $z=L$. With known $\hat a_1(0)$ and $\hat a_2(L)$, we rearrange Eq. \eqref{eq:SolutionBW1} and obtain solutions for $\hat a_1(L)$ and $\hat a_2(0)$:
\begin{widetext}
\begin{equation}
\begin{aligned}
\left[\begin{matrix}\hat a_{1}\left(L\right)\\\hat a_2^\dag\left(0\right)\\\end{matrix}\right]
=\left[\begin{matrix}A&B\\C&D\\\end{matrix}\right]\left[\begin{matrix}\hat a_{1}\left(0\right)\\\hat a_2^\dag\left(L\right)\\\end{matrix}\right]+\left[\begin{matrix}1&-B\\0&-D\\\end{matrix}\right]\int_{0}^{L}\left[\begin{matrix}\bar A_1\left(z\right)&\bar B_1\left(z\right)\\\bar C_1\left(z\right)&\bar D_1\left(z\right)\\\end{matrix}\right]\left(\mathrm{N_{BR}}\left[\begin{matrix}\hat f_{1}\left(z\right)\\\hat f_2^\dag\left(z\right)\\\end{matrix}\right]+\mathrm{N_{BI}}\left[\begin{matrix}\hat f_{1}^\dag\left(z\right)\\\hat f_2\left(z\right)\\\end{matrix}\right]\right)dz,
\end{aligned}\label{eq:SolutionBW2}
\end{equation}
\end{widetext}
~\\
~\\
where 
\begin{equation}
\begin{aligned}
&A=\bar A-\frac{\bar B\bar C}{\bar D},\\
&B=\frac{\bar B}{\bar D},\\
&C=-\frac{\bar C}{\bar D},\\
&D=\frac{1}{\bar D}.
\end{aligned}\label{barABCD}
\end{equation}
We numerically confirm that Eq. \eqref{eq:SolutionBW2} preserves the commutation relations 
\begin{equation}
\begin{aligned}
   \left[\hat a_1(\omega, L), \hat a_1^{\dag} (\omega', L)\right]=\left[\hat a_1(\omega, 0), \hat a_1^{\dag} (\omega', 0)\right],\\
   \left[\hat a_2(\omega, 0), \hat a_2^{\dag} (\omega', 0)\right]=\left[\hat a_2(\omega, L), \hat a_2^{\dag} (\omega', L)\right].
\end{aligned}
\label{eq:commutationBW}
\end{equation}

Similarly to the forward-wave configuration, we examine the following four special cases.

\noindent\textbf{Case 1:} We assume the nonlinear coupling coefficient $\kappa$ is real and both modes have losses ($\mathrm{Re}\{\alpha_m\}\geq 0$). Under such a condition, we have the following diagonalized noise matrix
\begin{equation}
\begin{aligned}
\mathrm{N_{B}}=\left[\begin{matrix}\sqrt{2\mathrm{Re}\{\alpha_1\}}&0\\0&-\sqrt{2\mathrm{Re}\{\alpha_2\}}\\\end{matrix}\right],
\end{aligned}\label{eq:NBCase1BW}
\end{equation}
and the coupled Langevin equations
\begin{equation}
\begin{aligned}
\frac{\partial}{\partial z}\left[\begin{matrix}\hat a_{1}\\\hat a_2^\dag\\\end{matrix}\right]
=\mathrm{M_{B}}\left[\begin{matrix}\hat a_{1}\\{\hat a_2^{\dag}}\\\end{matrix}\right]
+\left[\begin{matrix}\sqrt{2\mathrm{Re}\{\alpha_1\}}\hat f_{1}\\{-\sqrt{2\mathrm{Re}\{\alpha_2\}}\hat f_2^{\dag}}\\\end{matrix}\right].
\end{aligned}\label{eq:QLECase1BW}
\end{equation}

\noindent\textbf{Case 2:} $\kappa$ is real, mode 1 has linear loss ($\mathrm{Re}\{\alpha_1\}=\alpha\geq 0$), and mode 2 has linear gain ($\mathrm{Re}\{\alpha_2\}=-g\leq 0$). The noise matrix becomes
\begin{equation}
\begin{aligned}
\mathrm{N_{F}}=\left[\begin{matrix}\sqrt{2\alpha}&0\\0&-i\sqrt{2 g}\\\end{matrix}\right].
\end{aligned}\label{eq:NFCase1}
\end{equation}
We have the following coupled Langevin equations
\begin{equation}
\begin{aligned}
\frac{\partial}{\partial z}\left[\begin{matrix}\hat a_{1}\\\hat a_2^\dag\\\end{matrix}\right]
=\mathrm{M_{B}}\left[\begin{matrix}\hat a_{1}\\{\hat a_2^{\dag}}\\\end{matrix}\right]
+\left[\begin{matrix}\sqrt{2\alpha}\hat f_{1}\\-\sqrt{2 g}\hat f_2\\\end{matrix}\right].
\end{aligned}\label{eq:QLECase2BW}
\end{equation}

\noindent\textbf{Case 3:} The two modes are perfectly phase-matched without linear gain and loss: $\Delta k=0$, $\alpha_1=\alpha_2=0$, but the nonlinear coupling coefficient is complex-valued $\kappa=\eta+i\zeta$. In this case, the coupled matrix is
\begin{equation}
\begin{aligned}
\mathrm{M_{B}}=\left[\begin{matrix}0&-\zeta+i\eta\\-\zeta+i\eta&0\\\end{matrix}\right].
\end{aligned}\label{eq:MBWCase3}
\end{equation}
The noise matrix becomes 
\begin{equation}
\begin{aligned}
\mathrm{N_{B}}=\Theta(\zeta)\sqrt{\zeta}\left[\begin{matrix}1&1\\1&-1\\\end{matrix}\right]
+i\Theta(-\zeta)\sqrt{-\zeta}\left[\begin{matrix}1&1\\1&-1\\\end{matrix}\right].
\end{aligned}
\label{eq:NBCase3}
\end{equation}
The Langevin coupled equations are
\begin{equation}
\begin{aligned}
\frac{\partial}{\partial z}\left[\begin{matrix}\hat a_{1}\\\hat a_2^\dag\\\end{matrix}\right]
=&\mathrm{M_{B}}\left[\begin{matrix}\hat a_{1}\\{\hat a_2^{\dag}}\\\end{matrix}\right]
+\Theta(\zeta)\sqrt{\zeta}\left[\begin{matrix}1&1\\1&-1\\\end{matrix}\right]\left[\begin{matrix}\hat f_{1}\\{\hat f_2^{\dag}}\\\end{matrix}\right]\\
&+\Theta(-\zeta)\sqrt{-\zeta}\left[\begin{matrix}1&1\\1&-1\\\end{matrix}\right]\left[\begin{matrix}\hat f_{1}^{\dag}\\{\hat f_2}\\\end{matrix}\right].\\
\end{aligned}\label{eq:QLECase3BW}
\end{equation}
Eq. (\ref{eq:QLECase3BW}) shows that in the backward-wave configuration, a complex-valued nonlinear coupling coefficient also leads to Langevin noises even though there is no linear gain or loss.

\noindent\textbf{Case 4:} As $\kappa$ is real and there are equal losses in both modes ($\alpha_1=\alpha_2=\alpha>0$) with perfect phase matching ($\Delta k=0$), the coupled equations can be written as
\begin{equation}
\begin{aligned}
i\frac{\partial}{\partial z}\left[\begin{matrix}\hat a_{1}\\\hat a_2^\dag\\\end{matrix}\right]
=\left[\begin{matrix}-i\alpha&-\kappa\\-\kappa&i\alpha\\\end{matrix}\right]\left[\begin{matrix}\hat a_{1}\\{\hat a_2^{\dag}}\\\end{matrix}\right]=\hat{\mathcal{H}}\left[\begin{matrix}\hat a_{1}\\{\hat a_2^{\dag}}\\\end{matrix}\right].
\end{aligned}\label{eq:PTFWM}
\end{equation}
Interestingly, the effective Hamiltonian $\hat{\mathcal{H}}$ here follows parity-time (PT) symmetry \cite{PhysRevLett.80.5243, Miri_2019}. 

\section{Microscopic Origin of Langevin Noises: SFWM}\label{sec:Microscopic}

\begin{figure*}
\centering
\includegraphics[width=0.8\textwidth]{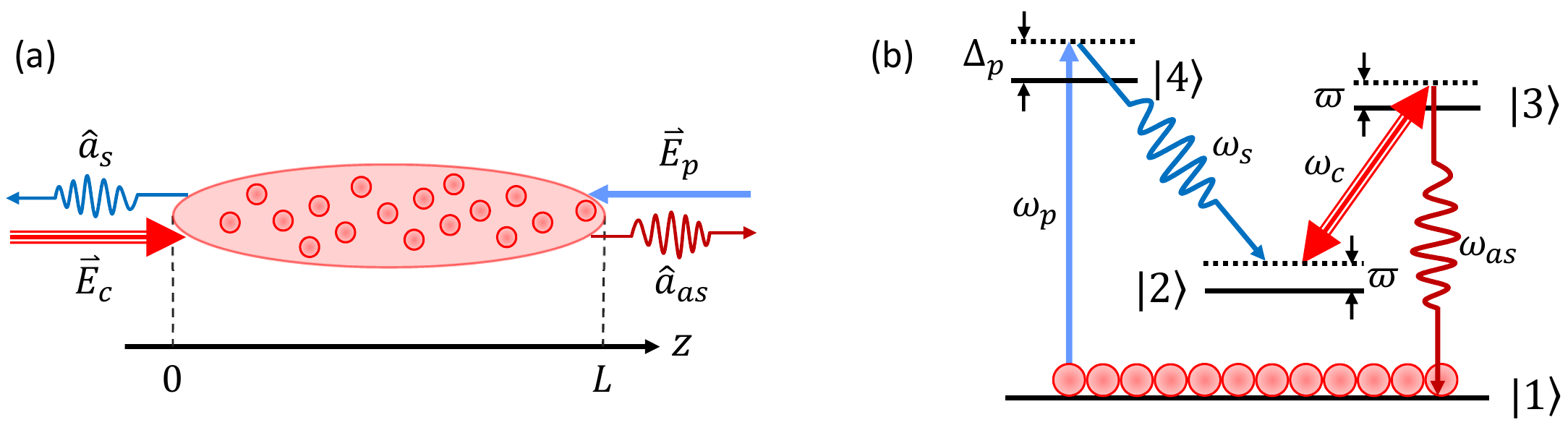}
\caption{Spontaneous four-wave mixing (SFWM) in a double-$\Lambda$ four-level cold atomic medium. (a) Backward-wave geometry of SFWM optical configuration. Driven by counter-propagating pump ($E_p$) and coupling ($E_c$) beams, phase-matched backward Stokes ($\hat a_s$) and anti-Stokes ($\hat a_{as}$) are spontaneously generated from a laser-cooled atomic medium.  (b) Atomic energy-level diagram. The pump ($\omega_p$) laser is detuned with $\Delta_p$ from transition $|1\rangle\rightarrow|4\rangle$, and the coupling ($\omega_c$) laser is on-resonant with transition $|2\rangle\rightarrow|3\rangle$. Stokes ($\omega_s$) photons are spontaneously generated from transition $|4\rangle\rightarrow|2\rangle$, and anti-Stokes ($\omega_{as}$) photons from transition $|3\rangle\rightarrow|1\rangle$. $\varpi=\omega_{as}-\omega_{13}$ is the anti-Stokes photon frequency detuning from transition $|1\rangle\rightarrow|3\rangle$.}
\label{fig:Fig02}
\end{figure*}

One could validate the above phenomenological approach of quantum Langevin coupled equations by confirming the microscopic origin of the Langevin noises. However, for two systems with the same quantum Langevin equations, their microscopic structures may be quite different. Therefore it is impossible to sort all microscopic systems. In this section, we focus on SFWM in a double-$\Lambda$ four-level atomic system \cite{PhysRevLett.93.183601, balic2005generation, kolchin2007electromagnetically, du2008narrowband, zhao2016narrowband} with electromagnetically induced transparency (EIT) \cite{EIT-Harris, RevModPhys.77.633}, and show that the phenomenological approach in the above section agrees with the numerical results from the microscopic quantum theory of light-atom interaction.

We start from a single-atom picture, considering an EM wave couples the atomic transition $\left|j\right>$ and $\left|k\right>$. The induced single atom polarization $\hat p_{jk}\propto \mu_{jk}\hat \sigma_{jk}$, where $\mu_{jk}$ is the electric dipole moment matrix element, $\hat \sigma_{jk}=\left|j\left>\right<k\right|$ is single atom transition operator from state $\left|k\right>$ to $\left|j\right>$. In the Heisenberg-Langevin picture, the single-atom transition operator can be expressed as 
\begin{equation}
\begin{aligned}
\hat \sigma_{jk}=\hat \sigma_{jk}^{(0)}+\sum_{\mu\nu}\beta_{\mu\nu}\hat f_{\mu\nu}^{(\sigma)},
\end{aligned}\label{eq:single atom}
\end{equation}
where $\hat \sigma_{jk}^{(0)}=\langle\hat\sigma_{jk}\rangle$ is the zeroth-order steady state solution. The single atom noise operator between atomic transition $\left|\nu\right> \rightarrow \left|\mu\right>$ is represented by $\hat f_{\mu\nu}^{(\sigma)}$, which satisfies the following correlations: 
\begin{equation}
\begin{aligned}
    \langle\hat f^{(\sigma)}_{\mu\nu}(\omega)\hat f^{(\sigma)\dag}_{\mu'\nu'}(\omega')\rangle=\langle\hat f^{(\sigma)}_{\mu\nu}(\omega)\hat f^{(\sigma)}_{\nu'\mu'}(\omega')\rangle \\ 
    =\mathcal{D}_{\mu\nu,\nu'\mu'}\delta(\omega-\omega'), \\
     \langle\hat f^{(\sigma)\dag}_{\mu\nu}(\omega)\hat f^{(\sigma)}_{\mu'\nu'}(\omega')\rangle=\langle\hat f^{(\sigma)}_{\nu\mu}(\omega)\hat f^{(\sigma)}_{\mu'\nu'}(\omega')\rangle \\ 
    =\mathcal{D}_{\nu\mu,\mu'\nu'}\delta(\omega-\omega'),
\end{aligned}
\label{eq:SingleAtomCorrelations}
\end{equation}
where $\mathcal{D}_{\mu\nu,\nu'\mu'}$ and $\mathcal{D}_{\nu\mu,\mu'\nu'}$ are diffusion coefficients. 

In a continuous medium with atomic number density $n$, the noises from different atoms are uncorrelated. We have the spatially averaged atomic operator 
\begin{equation}
\begin{aligned}
{\hat{\bar\sigma}}_{jk}\equiv{\hat{\sigma}}_{jk}^{(0)}+\frac{1}{\sqrt{nA}}\sum_{\mu\nu}\beta_{\mu\nu}\hat{\bar{f}}^{(\sigma)}_{\mu\nu},
\end{aligned}\label{eq:average sigma}
\end{equation}
where $A$ is the single-mode cross-section area, and the spatially averaged atomic noise operators $\hat{\bar{f}}^{(\sigma)}_{\mu\nu}$
satisfy the following modified correlations
\begin{equation}
\begin{aligned}
    \langle\hat{\bar{f}}^{(\sigma)}_{\mu\nu}(\omega,z)\hat{\bar{f}}^{(\sigma)\dag}_{\mu'\nu'}(\omega',z')\rangle=\langle\hat{\bar{f}}^{(\sigma)}_{\mu\nu}(\omega,z)\hat{\bar{f}}^{(\sigma)}_{\nu'\mu'}(\omega',z')\rangle\\ 
    =\mathcal{D}_{\mu\nu,\nu'\mu'}\delta(\omega-\omega')\delta(z-z'), \\
     \langle\hat{\bar{f}}^{(\sigma)\dag}_{\mu\nu}(\omega,z)\hat{\bar{f}}^{(\sigma)}_{\mu'\nu'}(\omega',z')\rangle=\langle\hat{\bar{f}}^{(\sigma)}_{\nu\mu}(\omega,z)\hat{\bar{f}}^{(\sigma)}_{\mu'\nu'}(\omega',z')\rangle \\ 
    =\mathcal{D}_{\nu\mu,\mu'\nu'}\delta(\omega-\omega')\delta(z-z'),
\end{aligned}
\label{eq:AveragedAtomCorrelations}
\end{equation}
where the diffusion coefficients are the same as those from the single-atom picture.

The electric field and polarization are described as
\begin{equation}
\begin{aligned}
&\hat{\mathbf{E}}(t,z)=\frac{1}{2}\left[\hat E^{(+)}(t,z)+\hat E^{(-)}(t,z)\right],\\
&\hat{\mathbf{P}}(t,z)=\frac{1}{2}\left[\hat P^{(+)}(t,z)+\hat P^{(-)}(t,z)\right],
\end{aligned}\label{eq:SVEA1}
\end{equation}
Where $\hat E^{(+)}, \hat P^{(+)}$ and $\hat E^{(-)}, \hat P^{(-)}$ are positive and negative frequency parts. We take the following Fourier transform
\begin{equation}
\begin{aligned}
&\hat E^{(+)}(t,z)=\frac{1}{\sqrt{2\pi}}\int d\omega \hat E(\omega,z)e^{i\left(\pm \frac{\omega}{c}z-\omega t\right)},\\
&\hat P^{(+)}(t,z)=\frac{1}{\sqrt{2\pi}}\int d\omega \hat P(\omega,z)e^{i\left(\pm \frac{\omega}{c}z-\omega t\right)},
\end{aligned}\label{eq:E+}
\end{equation}
where $\hat E(\omega,z), \hat P(\omega,z)$ are complex amplitudes in frequency domain. The Maxwell equation under slowly varying envelope approximation (SVEA) can be written as
\begin{equation}
\begin{aligned}
\pm\frac{\partial \hat E(\omega,z)}{\partial z}=\frac{i}{2}\omega\eta \hat P(\omega,z),
\end{aligned}\label{eq:SVEA}
\end{equation}
where $\pm$ represents for propagation direction along $\pm z$, and free space impedance $\eta=1/(c\varepsilon_0)=377$ Ohm, with $c$ being the speed of light in vacuum, and $\varepsilon_0$ the vacuum permittivity. With quantized electric field
\begin{equation}
\begin{aligned}
\hat E(\omega,z)=\sqrt{\frac{2\hbar\omega}{c\varepsilon_0A}}\hat{a}(\omega,z),
\end{aligned}\label{eq:E}
\end{equation}
and
\begin{equation}
\begin{aligned}
\hat P(\omega,z)=2n\mu_{jk}{\hat{\bar\sigma}}_{jk}(\omega,z),
\end{aligned}\label{eq:P}
\end{equation}
we obtain the Langevin equation for the EM field in the atomic medium 
\begin{equation}
\begin{aligned}
\pm\frac{\partial\hat{a}(\omega,z)}{\partial z}&=i\,nAg_{jk}{\hat{\bar\sigma}}_{jk}(\omega,z)\\
&=i\,nAg_{jk}{\hat{\sigma}}_{jk}^{(0)}(\omega,z)+\hat {\bar F}(\omega,z),
\end{aligned}\label{Quantized Maxwell}
\end{equation}
where 
\begin{equation}
\begin{aligned}
&g_{jk}=\mu_{jk}\sqrt{\frac{\omega_{jk}}{2 c\varepsilon_0\hbar A}}, \\
&\hat{\bar F}(\omega,z)=i\sqrt{nA}g_{jk}\sum_{\mu\nu}\beta_{\mu\nu}\hat{\bar{f}}^{(\sigma)}_{\mu\nu}(\omega,z)\\
&=i\mu_{jk}\sqrt{\frac{n\omega_{jk}}{2c\varepsilon_0\hbar}}\sum_{\mu\nu}\beta_{\mu\nu}\hat{\bar{f}}^{(\sigma)}_{\mu\nu}(\omega,z).
\end{aligned}\label{eq:F_def}
\end{equation}
Here $g_{jk}=g_{kj}^*$ is single photon-atom coupling strength.

Now we turn to the backward-wave SFWM in a double-$\Lambda$ four-level atomic system as illustrated in Fig. \ref{fig:Fig02}. In presence of counter-propagating pump ($E_p, \omega_p$) and coupling ($E_c, \omega_c$) laser beams, phase-matched Stokes ($\omega_s$) and anti-Stokes ($\omega_{as}$) are spontaneously generated and propagate through the medium in opposing directions. In the rotating reference frame, the interaction Hamiltonian for a single atom is
\begin{eqnarray}
\hat{V}=&-&\hbar\left(g_{31}{\hat{a}}_{as}{\hat{\sigma}}_{31}+g_{13}{\hat{a}}_{as}^\dag{\hat{\sigma}}_{13}\right)-\hbar\left(g_{42}{\hat{a}}_s{\hat{\sigma}}_{42}+g_{24}{\hat{a}}_s^\dag{\hat{\sigma}}_{24}\right)\nonumber\\
&-&\frac{1}{2}\hbar\left(\Omega_c{\hat{\sigma}}_{32}+\Omega_c^\ast{\hat{\sigma}}_{23}\right)-\frac{1}{2}\hbar\left(\Omega_p{\hat{\sigma}}_{41}+\Omega_p^\ast{\hat{\sigma}}_{14}\right)\nonumber\\
&-&\hbar\Delta_p{\hat{\sigma}}_{44}-\hbar\varpi{\hat{\sigma}}_{33}-\hbar\varpi{\hat{\sigma}}_{22},
\end{eqnarray}
where $\Omega_c=\mu_{32}E_c/\hbar$ is coupling Rabi frequency. The coupling laser is on-resonant with transition $\left|2\right\rangle\rightarrow\left|3\right\rangle$. $\Omega_p=\mu_{41}E_p/\hbar$ is pump Rabi frequency. The pump laser is far detuned from the transition $\left|1\right\rangle\rightarrow\left|4\right\rangle$ with $\Delta_p=\omega_p-\omega_{14}$ so that the atomic population mainly occupies the ground state $\left|1\right\rangle$. We take this ground-state approximation through this section. With continuous-wave pump and coupling driving fields, the energy conservation leads to $\omega_{as}+\omega_{s}=\omega_c+\omega_p$. Here $\varpi=\omega_{as}-\omega_{13}$ is the anti-Stokes frequency detuning and thus the Stokes frequency detuning is $\omega_{s}-\omega_{s0}=-\varpi$. 

The atomic evolution is governed by the following Heisenberg-Langevin equation~\cite{kolchin2007electromagnetically}
\begin{equation}
\begin{aligned}
\frac{\partial}{\partial t}\hat{\sigma}_{jk}=\frac{i}{\hbar}[\hat V,\hat\sigma_{jk}]-\gamma_{jk}\hat\sigma_{jk}+r_{jk}^A+\hat f_{jk}^{(\sigma)},
\end{aligned}\label{eq:HEM}
\end{equation}
where $\gamma_{jk}=\gamma_{kj}$ (nonzero only as $j\neq k$) are dephasing rates, $r_{jk}^A$ (nonzero only as $j=k$) are the population transfer resulting from spontaneous emission decay. The full equation of motion can be found in Appendix \ref{Appendix: Heisenberg-Langevin Equations in FWM}. The diffusion coefficients $\mathcal{D}_{jk,j^\prime k^\prime}$ can be obtained through the Einstein relation
\begin{equation}
\begin{aligned}
\mathcal{D}_{jk,j^\prime k^\prime}
&=\frac{\partial}{\partial t}\left<{\hat{\sigma}}_{jk}{\hat{\sigma}}_{j^\prime k^\prime}\right>\\
&-\left<{\hat{A}}_{jk}{\hat{\sigma}}_{j^\prime k^\prime}\right>-\left<{\hat{\sigma}}_{jk}{\hat{A}}_{j^\prime k^\prime}\right>,
\end{aligned}\label{eq:TwoLevel Dxy}
\end{equation}
where ${\hat{A}}_{jk}=\frac{\partial}{\partial t}{\hat{\sigma}}_{jk}-{\hat{f}}_{jk}^{(\sigma)}$. For the SFWM governed by Eq.~\eqref{eq:HEM}, we have \cite{kolchin2007electromagnetically, zhao2016narrowband}
\begin{widetext}
\begin{equation}
\begin{aligned}
&\left[\begin{matrix}\begin{matrix}\mathcal{D}_{12,21}&\mathcal{D}_{12,24}\\\mathcal{D}_{42,21}&\mathcal{D}_{42,24}\\\end{matrix}&\begin{matrix}\mathcal{D}_{12,31}&\mathcal{D}_{12,34}\\\mathcal{D}_{42,31}&\mathcal{D}_{42,34}\\\end{matrix}\\\begin{matrix}\mathcal{D}_{13,21}&\mathcal{D}_{13,24}\\\mathcal{D}_{43,21}&\mathcal{D}_{43,24}\\\end{matrix}&\begin{matrix}\mathcal{D}_{13,31}&\mathcal{D}_{13,34}\\\mathcal{D}_{43,31}&\mathcal{D}_{43,34}\\\end{matrix}\\\end{matrix}\right]\\
&=\left[\begin{matrix}2\gamma_{12}\left<{\hat{\sigma}}_{11}\right>+\Gamma_{31}\left<{\hat{\sigma}}_{33}\right>+\Gamma_{41}\left<{\hat{\sigma}}_{44}\right>&\gamma_{12}\left<{\hat{\sigma}}_{14}\right>&0&0
\\\gamma_{12}\left<{\hat{\sigma}}_{41}\right>&0&0&0
\\0&0&\Gamma_{3}\left<{\hat{\sigma}}_{11}\right>+\Gamma_{31}\left<{\hat{\sigma}}_{33}\right>+\Gamma_{41}\left<{\hat{\sigma}}_{44}\right>&\Gamma_3\left<{\hat{\sigma}}_{14}\right>
\\0&0&\Gamma_3\left<{\hat{\sigma}}_{41}\right>&\Gamma_3\left<{\hat{\sigma}}_{44}\right>\end{matrix}\right],
\end{aligned}\label{eq:FWM Diffusion}
\end{equation}
\begin{equation}
\begin{aligned}
&\left[\begin{matrix}\begin{matrix}\mathcal{D}_{21,12}&\mathcal{D}_{21,42}\\\mathcal{D}_{24,12}&\mathcal{D}_{24,42}\\\end{matrix}&\begin{matrix}\mathcal{D}_{21,13}&\mathcal{D}_{21,43}\\\mathcal{D}_{24,13}&\mathcal{D}_{24,43}\\\end{matrix}\\\begin{matrix}\mathcal{D}_{31,12}&\mathcal{D}_{31,42}\\\mathcal{D}_{34,12}&\mathcal{D}_{34,42}\\\end{matrix}&\begin{matrix}\mathcal{D}_{31,13}&\mathcal{D}_{31,43}\\\mathcal{D}_{34,13}&\mathcal{D}_{34,43}\\\end{matrix}\\\end{matrix}\right]\\
&=\left[\begin{matrix}2\gamma_{12}\left<{\hat{\sigma}}_{22}\right>+\Gamma_{32}\left<{\hat{\sigma}}_{33}\right>+\Gamma_{42}\left<{\hat{\sigma}}_{44}\right>&0&\gamma_{12}\left<{\hat{\sigma}}_{23}\right>&0
\\0&\Gamma_{4}\left<{\hat{\sigma}}_{22}\right>+\Gamma_{32}\left<{\hat{\sigma}}_{33}\right>+\Gamma_{42}\left<{\hat{\sigma}}_{44}\right>&0&\Gamma_{4}\left<{\hat{\sigma}}_{23}\right>
\\\gamma_{12}\left<{\hat{\sigma}}_{32}\right>&0&0&0
\\0&\Gamma_4\left<{\hat{\sigma}}_{32}\right>&0&\Gamma_4\left<{\hat{\sigma}}_{33}\right>\end{matrix}\right].
\end{aligned}\label{eq:FWM Diffusion 1}
\end{equation}
\end{widetext}

\begin{figure*} 
\centering
\includegraphics[width=0.85\textwidth]{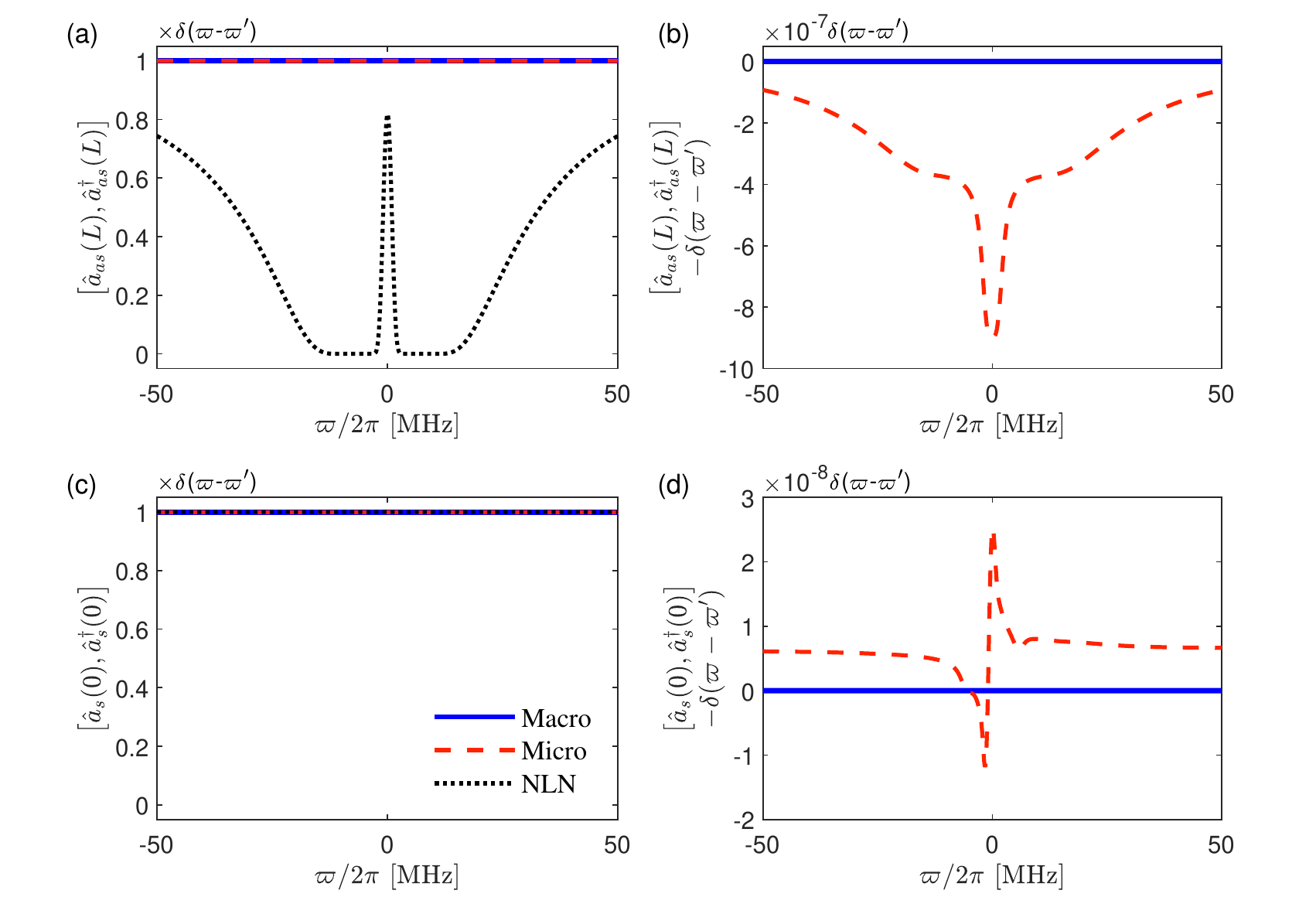}
\caption{Comparison of commutation relations between the macroscopic (``Macro'', blue solid lines) and microscopic (``Micro'', red dashed lines) approaches in the group delay regime: (a) $[\hat a_{as}(L), \hat a_{as}^{\dag}(L)]$, (b) $[\hat a_{as}(L), \hat a_{as}^{\dag}(L)]-\delta(\varpi-\varpi')$, (c) $[\hat a_{s}(0), \hat a_{s}^{\dag}(0)]$, and (d)$[\hat a_{s}(0), \hat a_{s}^{\dag}(0)]-\delta(\varpi-\varpi')$. The results with no Langevin noise operators (``NLN'') are shown as black dotted lines in (a) and (c).} 
\label{fig:Fig03}
\end{figure*}

\begin{figure*} 
\centering
\includegraphics[width=0.85\textwidth]{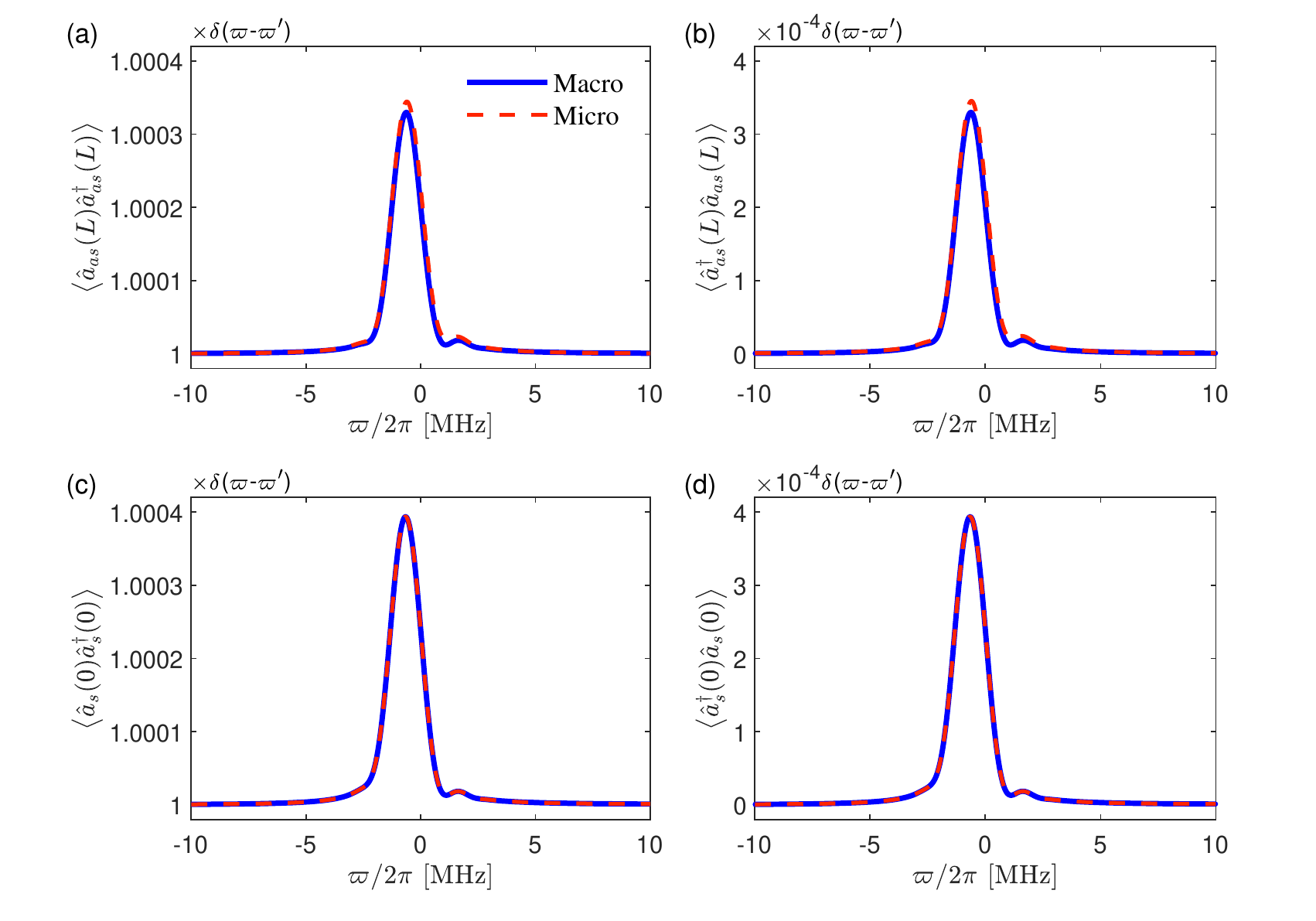}
\caption{Four real correlations of Stokes and anti-Stokes fields in the group delay regime: (a) $\langle \hat a_{as}(L)\hat a_{as}^{\dag}(L)\rangle$, (b) $\langle \hat a_{as}^{\dag}(L)\hat a_{as}(L)\rangle$, (c) $\langle \hat a_{s}(0)\hat a_{s}^{\dag}(0)\rangle$, and (d) $\langle \hat a_{s}^{\dag}(0)\hat a_{s}(0)\rangle$. The macroscopic (``Macro'') and microscopic (``Micro'') approaches are shown as blue solid and red dashed lines, respectively.}
\label{fig:Fig04}
\end{figure*}

\begin{figure*} 
\centering
\includegraphics[width=0.85\textwidth]{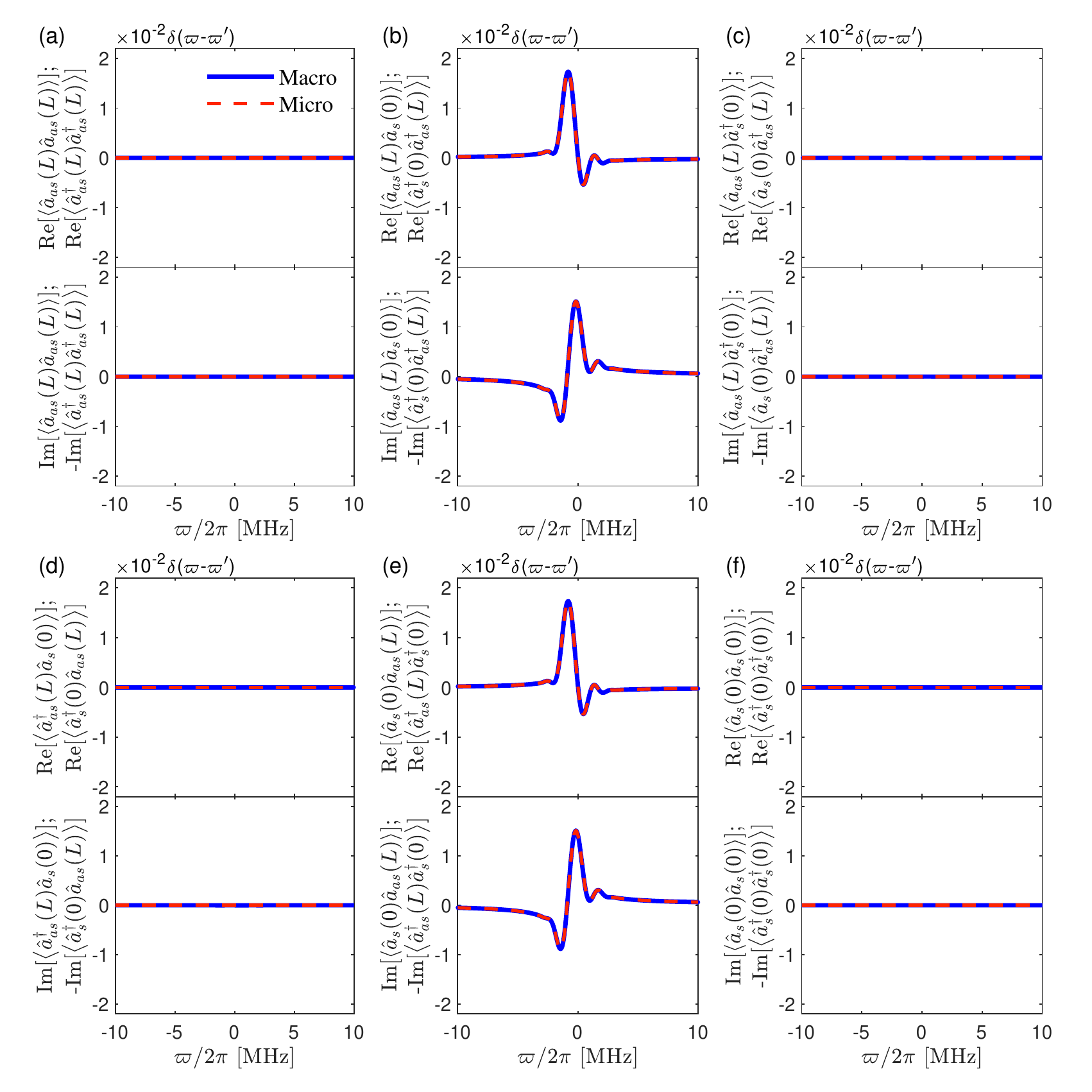}
\caption{Twelve complex correlations of Stokes and anti-Stokes fields in the group delay regime: (a) $\langle \hat a_{as}(L)\hat a_{as}(L)\rangle=\langle \hat a_{as}^{\dag}(L)\hat a_{as}^{\dag}(L)\rangle^*$, (b) $\langle \hat a_{as}(L)\hat a_{s}(0)\rangle=\langle \hat a_{s}^{\dag}(0)\hat a_{as}^{\dag}(L)\rangle^*$, (c) $\langle \hat a_{as}(L)\hat a_{s}^{\dag}(0)\rangle=\langle \hat a_{s}(0)\hat a_{as}^{\dag}(L)\rangle^*$, (d) $\langle \hat a_{as}^{\dag}(L)\hat a_{s}(0)\rangle=\langle \hat a_{s}^{\dag}(0)\hat a_{as}(L)\rangle^*$, (e) $\langle \hat a_{s}(0)\hat a_{as}(L)\rangle=\langle \hat a_{as}^{\dag}(L)\hat a_{s}^{\dag}(0)\rangle^*$, and (f) $\langle \hat a_{s}(0)\hat a_{s}(0)\rangle=\langle \hat a_{s}^{\dag}(0)\hat a_{s}^{\dag}(0)\rangle^*$. The macroscopic (``Macro'') and microscopic (``Micro'') approaches are shown as blue solid and red dashed lines, respectively.}
\label{fig:Fig05}
\end{figure*}

\begin{figure*} 
\centering
\includegraphics[width=0.85\textwidth]{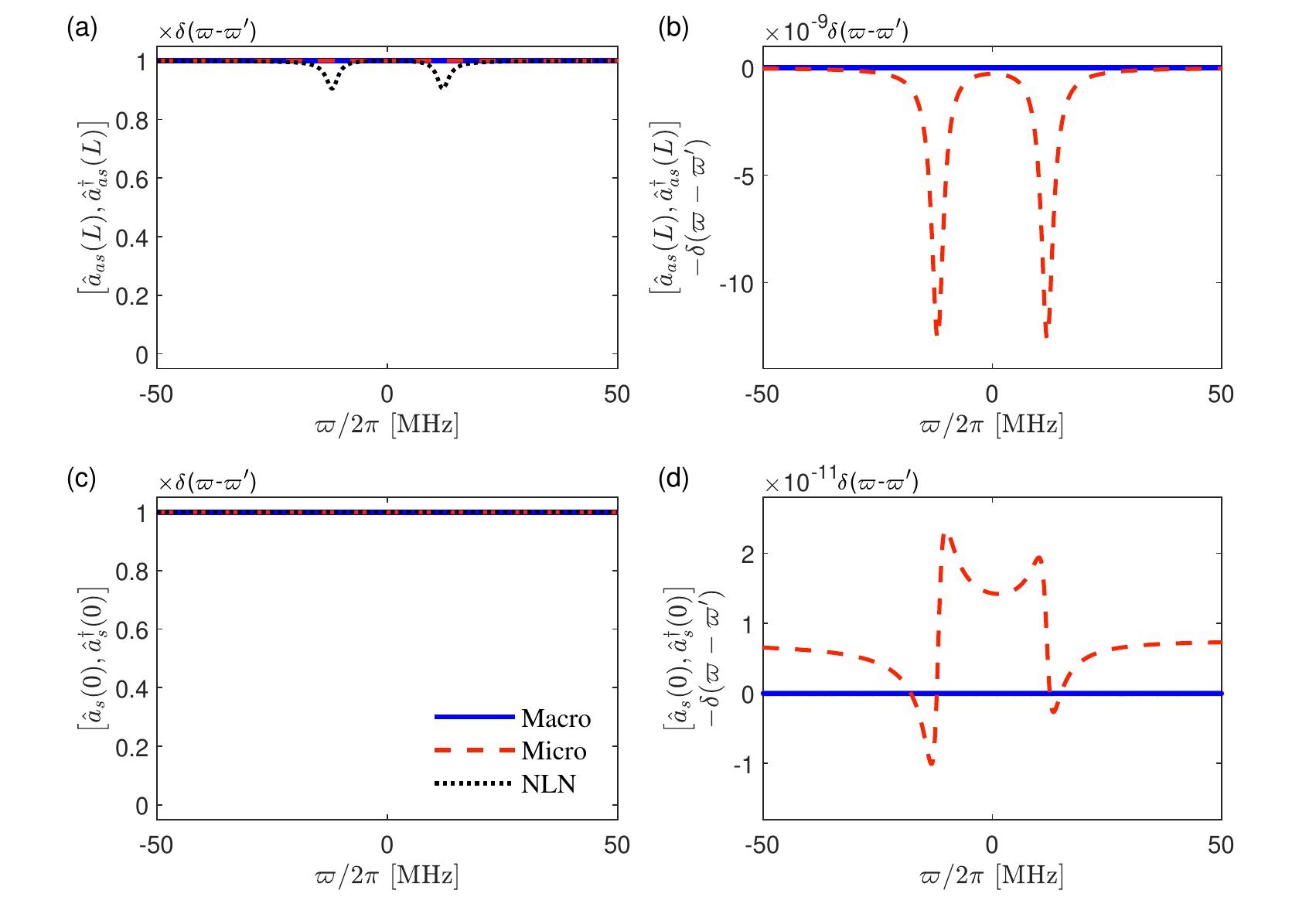}
\caption{Comparison of commutation relations between the macroscopic (``Macro'', blue solid lines) and microscopic (``Micro'', red dashed lines) approaches in  the damped Rabi oscillation regime: (a) $[\hat a_{as}(L), \hat a_{as}^{\dag}(L)]$, (b) $[\hat a_{as}(L), \hat a_{as}^{\dag}(L)]-\delta(\varpi-\varpi')$, (c) $[\hat a_{s}(0), \hat a_{s}^{\dag}(0)]$, and (d)$[\hat a_{s}(0), \hat a_{s}^{\dag}(0)]-\delta(\varpi-\varpi')$. The results with no Langevin noise operators (``NLN'') are shown as black dotted lines in (a) and (c).} 
\label{fig:Fig06}
\end{figure*}

\begin{figure*} 
\centering
\includegraphics[width=0.85\textwidth]{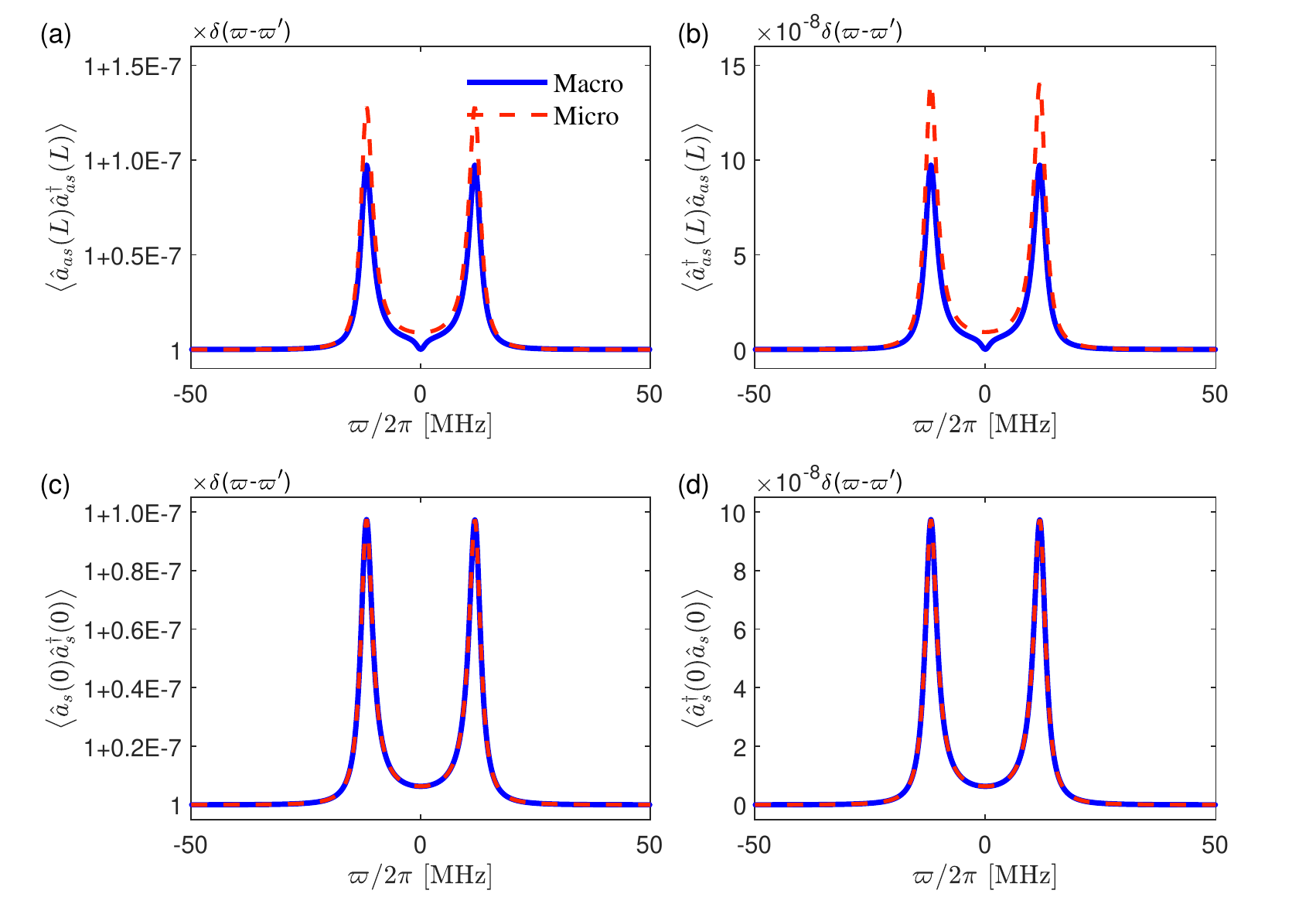}
\caption{Four real correlations of Stokes and anti-Stokes fields in the damped Rabi oscillation regime: (a) $\langle \hat a_{as}(L)\hat a_{as}^{\dag}(L)\rangle$, (b) $\langle \hat a_{as}^{\dag}(L)\hat a_{as}(L)\rangle$, (c) $\langle \hat a_{s}(0)\hat a_{s}^{\dag}(0)\rangle$, and (d) $\langle \hat a_{s}^{\dag}(0)\hat a_{s}(0)\rangle$. The macroscopic (``Macro'') and microscopic (``Micro'') approaches are shown as blue solid and red dashed lines, respectively.}
\label{fig:Fig07}
\end{figure*}

\begin{figure*} 
\centering
\includegraphics[width=0.85\textwidth]{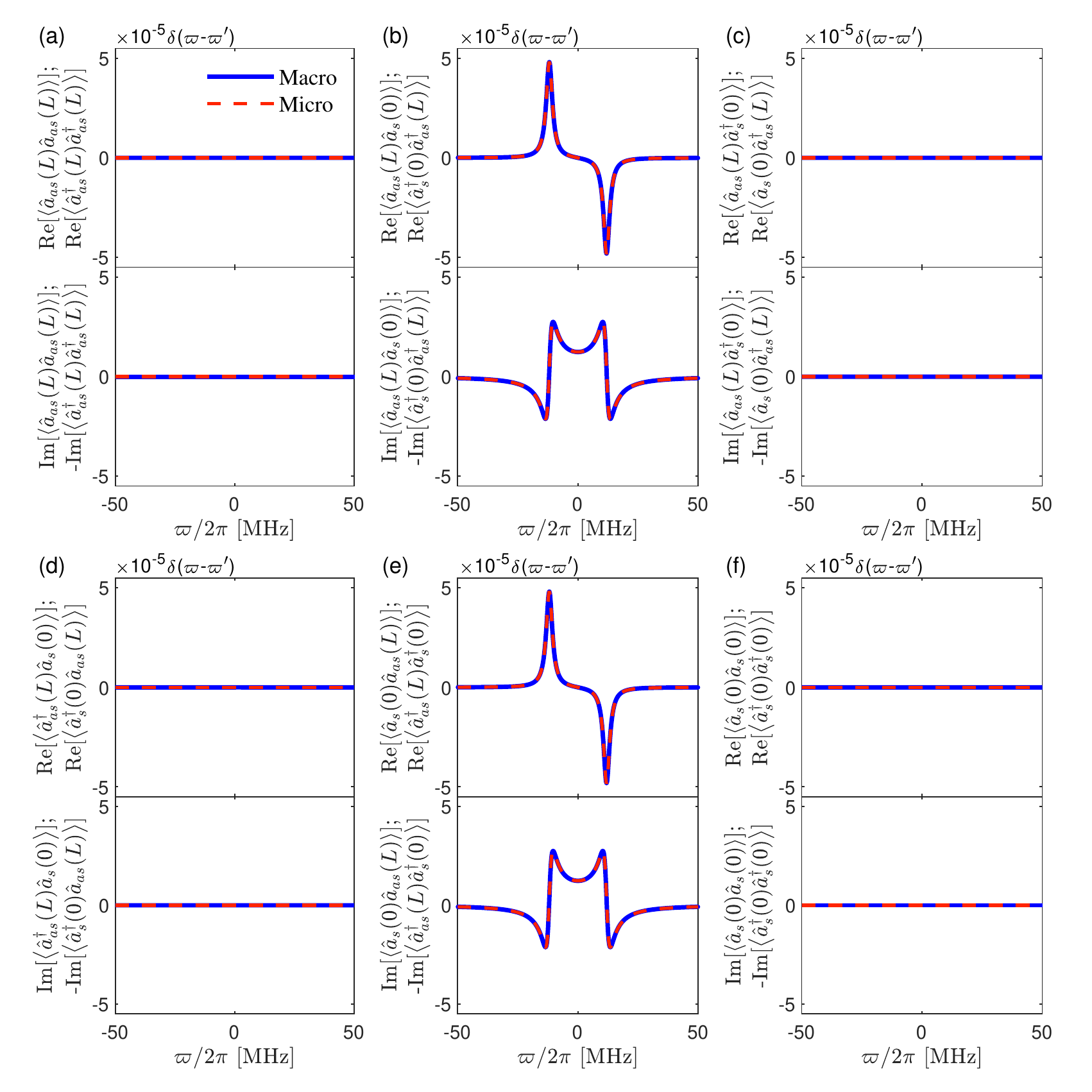}
\caption{Twelve complex correlations of Stokes and anti-Stokes fields in the damped Rabi oscillation regime: (a) $\langle \hat a_{as}(L)\hat a_{as}(L)\rangle=\langle \hat a_{as}^{\dag}(L)\hat a_{as}^{\dag}(L)\rangle^*$, (b) $\langle \hat a_{as}(L)\hat a_{s}(0)\rangle=\langle \hat a_{s}^{\dag}(0)\hat a_{as}^{\dag}(L)\rangle^*$, (c) $\langle \hat a_{as}(L)\hat a_{s}^{\dag}(0)\rangle=\langle \hat a_{s}(0)\hat a_{as}^{\dag}(L)\rangle^*$, (d) $\langle \hat a_{as}^{\dag}(L)\hat a_{s}(0)\rangle=\langle \hat a_{s}^{\dag}(0)\hat a_{as}(L)\rangle^*$, (e) $\langle \hat a_{s}(0)\hat a_{as}(L)\rangle=\langle \hat a_{as}^{\dag}(L)\hat a_{s}^{\dag}(0)\rangle^*$, and (f) $\langle \hat a_{s}(0)\hat a_{s}(0)\rangle=\langle \hat a_{s}^{\dag}(0)\hat a_{s}^{\dag}(0)\rangle^*$. The macroscopic (``Macro'') and microscopic (``Micro'') approaches are shown as blue solid and red dashed lines, respectively.}
\label{fig:Fig08}
\end{figure*}

Solving Eq.~\eqref{eq:HEM} under the ground-state approximation $\left<\hat{\sigma}_{11}\right>\cong 1$ with weak pump excitation $\Delta_p\gg\{\Omega_p, \Gamma_4\}$, we get the single-atom steady-state solutions (with $\mu\nu=12, 13, 42, 43$)
\begin{equation}
\begin{aligned}
&{\hat{\sigma}}_{13}=\hat\sigma_{13}^{(0)}+\sum_{\mu\nu}\beta_{\mu\nu}^{as}\hat f^{(\sigma)}_{\mu\nu},\\
&{\hat{\sigma}}_{42}=\hat\sigma_{42}^{(0)}+\sum_{\mu\nu}\beta_{\mu\nu}^{s}\hat f^{(\sigma)}_{\mu\nu},
\end{aligned}
\end{equation}
where
\begin{equation}
\begin{aligned}
\hat{\sigma}_{13}^{(0)}=&\frac{4\left(\varpi+i\gamma_{12}\right)}{T\left(\varpi\right)}g_{31}{\hat{a}}_{as}\\
&+\frac{\Omega_c\Omega_p}{T\left(\varpi\right)\left(\Delta_p+i\gamma_{14}\right)}g_{24}{\hat{a}}_s^\dag,\\
\hat{\sigma}_{42}^{(0)}=&\frac{\left(\varpi+i\gamma_{13}\right)}{T\left(\varpi\right)}\frac{\left|\Omega_p\right|^2}{\left(\Delta_p-i\gamma_{24}\right)}\frac{1}{\left(\Delta_p+i\gamma_{14}\right)}g_{24}{\hat{a}}_s^\dag\\
&+\frac{\Omega_p^\ast\Omega_c^\ast}{T\left(\varpi\right)\left(\Delta_p-i\gamma_{24}\right)}g_{31}{\hat{a}}_{as},
\end{aligned}
\label{eq:beta as}
\end{equation}
\begin{equation}
\begin{aligned}
&\beta_{12}^{as}=\frac{i2\Omega_c}{T\left(\varpi\right)},\\
&\beta_{13}^{as}=-\frac{i4\left(\varpi+i\gamma_{12}\right)}{T\left(\varpi\right)},\\
&\beta_{42}^{as}=-\frac{i\Omega_c\Omega_p}{T\left(\varpi\right)\left(\Delta_p-i\gamma_{24}\right)},\\
&\beta_{43}^{as}=\frac{{i2\Omega}_p\left(\varpi+i\gamma_{12}\right)}{T\left(\varpi\right)\left(\Delta_p-i\gamma_{34}\right)},\\
&\beta_{12}^{s}=\frac{i2\left(\varpi+i\gamma_{13}\right)}{T\left(\varpi\right)}\frac{\Omega_p^\ast}{\left(\Delta_p-i\gamma_{24}\right)},\\
&\beta_{13}^{s}=-\frac{i\Omega_p^\ast\Omega_c^\ast}{T\left(\varpi\right)\left(\Delta_p-i\gamma_{24}\right)},\\
&\beta_{42}^{s}=-\frac{i}{\left(\Delta_p-i\gamma_{24}\right)},\\
&\beta_{43}^{s}=-\frac{i\Omega_c^\ast}{2\left(\Delta_p-i\gamma_{24}\right)\left(\Delta_p-i\gamma_{34}\right)},
\end{aligned}\label{eq:beta s}
\end{equation}
where $T(\varpi)\equiv\left|\Omega_c\right|^2-4\left(\varpi+i\gamma_{13}\right)\left(\varpi+i\gamma_{12}\right)$. We then obtain the ensemble spatially averaged atomic operators for generating anti-Stokes and Stokes fields from Eq. \eqref{eq:average sigma} 
\begin{equation}
\begin{aligned}
&{\hat{\bar \sigma}}_{13}=\hat\sigma_{13}^{(0)}+\frac{1}{\sqrt{nA}}\sum_{\mu\nu}\beta_{\mu\nu}^{as}{\hat{\bar f}}^{(\sigma)}_{\mu\nu},\\
&{\hat{\bar \sigma}}_{42}=\hat\sigma_{42}^{(0)}+\frac{1}{\sqrt{nA}}\sum_{\mu\nu}\beta_{\mu\nu}^{s}{\hat{\bar f}}^{(\sigma)}_{\mu\nu}.
\end{aligned}\label{eq:FWMsigma}
\end{equation}
Following the procedures in Eqs. \eqref{Quantized Maxwell} and \eqref{eq:F_def}, 
\begin{equation}
\begin{aligned}
\frac{\partial\hat{a}_{as}(\omega,z)}{\partial z}&=i\,nAg_{13}{\hat{\bar\sigma}}_{13}(\omega,z),\\
\frac{\partial\hat{a}_{s}^\dag(\omega,z)}{\partial z}&=i\,nAg_{42}{\hat{\bar\sigma}}_{42}(\omega,z),
\end{aligned}\label{AS-S Maxwell}
\end{equation}
we get coupled equations for counter-propagating anti-Stokes (propagating along $+z$) and Stokes (propagating along $-z$) fields in the backward-wave configuration 
\begin{equation}
\begin{aligned}
\frac{\partial}{\partial z}\left[\begin{matrix}\hat a_{as}\\\hat a_s^\dag\\\end{matrix}\right]
=\left[\begin{matrix}-\alpha_{as}+i\frac{\Delta k}{2}&i\kappa_{as}\\i\kappa_{s}&\alpha_s^*-i\frac{\Delta k}{2}\\\end{matrix}\right]\left[\begin{matrix}\hat a_{as}\\{\hat a_s^{\dag}}\\\end{matrix}\right]+\left[\begin{matrix}\hat{\bar F}_{as}\\ -\hat {\bar F}_s^{\dag}\\\end{matrix}\right],
\end{aligned}\label{eq:FWM1}
\end{equation}
where
\begin{equation}
\begin{aligned}
&\hat{\bar F}_{as}=
ig_{13}\sqrt{nA}\left[\beta_{12}^{as}\hat{\bar f}^{(\sigma)}_{12}
+\beta_{13}^{as}\hat{\bar f}^{(\sigma)}_{13}
+\beta_{42}^{as}\hat{\bar f}^{(\sigma)}_{42}
+\beta_{43}^{as}\hat{\bar f}^{(\sigma)}_{43}\right],\\
&\hat{\bar F}_{s}^\dag=-
ig_{42}\sqrt{nA}\left[\beta_{12}^s\hat{\bar f}^{(\sigma)}_{12}
+\beta_{13}^s\hat{\bar f}^{(\sigma)}_{13}
+\beta_{42}^s\hat{\bar f}^{(\sigma)}_{42}
+\beta_{43}^s\hat{\bar f}^{(\sigma)}_{43}\right],
\end{aligned}\label{eq:F}
\end{equation}
and
\begin{equation}
\begin{aligned}
&\alpha_{as}=-i\frac{\omega_{as}}{2c}\chi_{as},\\
&\alpha_s=-i\frac{\omega_s}{2c}\ \chi_s,\\
&\kappa_{as}=\frac{\sqrt{\omega_{as}\omega_s}}{2c}\chi_{as}^{\left(3\right)}E_pE_c,\\
&\kappa_s=\frac{\sqrt{\omega_s\omega_{as}}}{2c}\chi_s^{\left(3\right)\ast}E_p^\ast E_c^\ast,\\
&\chi_{as}=\frac{4n\left|\mu_{13}\right|^2}{\varepsilon_0\hbar}\frac{\left(\varpi+i\gamma_{12}\right)}{T\left(\varpi\right)},\\
&\chi_s=\frac{n\left|\mu_{24}\right|^2}{\varepsilon_0\hbar}\frac{\left(\varpi-i\gamma_{13}\right)}{T^*\left(\varpi\right)}\frac{\left|\Omega_p\right|^2}{\Delta_p^2+\gamma_{14}^2},\\
&\chi_{as}^{\left(3\right)}=\frac{{n\mu}_{13}\mu_{32}\mu_{24}\mu_{41}}{\varepsilon_0\hbar^3}\frac{1}{T\left(\varpi\right)}\frac{1}{\left(\Delta_p+i\gamma_{14}\right)},\\
&\chi_s^{\left(3\right)}=\frac{n\mu_{13}\mu_{32}\mu_{24}\mu_{41}}{\varepsilon_0\hbar^3}\frac{1}{T^*\left(\varpi\right)}\frac{1}{\left(\Delta_p+i\gamma_{14}\right)},
\end{aligned}\label{eq:parameters}
\end{equation}
The expressions for $\beta_{\mu\nu}^{as}$ and $\beta_{\mu\nu}^s$ are listed in Eqs. (\ref{eq:beta s}). $\Delta k=(\omega_{as}-\omega_{s})/c-(\vec{k}_c+\vec{k}_p)\cdot \hat{z}$ is the phase mismatching in vacuum. Here the complex $\alpha_{as}$ represents the EIT loss and phase dispersion. $\alpha_{s}^{*}$ is the Raman gain and dispersion along $-z$ propagation direction. One can show that the nonlinear coupling coefficients can be expressed as $\kappa_{as}=\kappa e^{i\theta}$ and $\kappa_{s}=\kappa e^{-i\theta}$, where 
\begin{equation}
\begin{aligned}
\kappa=\frac{\sqrt{\omega_{as}\omega_s}}{2c}\frac{{n\mu}_{13}\mu_{24}}{\varepsilon_0\hbar}\left|\frac{\Omega_p \Omega_c}{\Delta_p+i\gamma_{14}}\right|\frac{1}{T(\varpi)},
\end{aligned}\label{eq:kappa}
\end{equation}
and $\theta$ is the phase of $\Omega_p \Omega_c/(\Delta_p+i\gamma_{14})$. As a result, $\kappa_{as}$ and $\kappa_{s}$ fulfill the gauge transformation discussed in Sec. \ref{sec:QLE}. Therefore, to be consistent with the treatment in Sec. \ref{sec:QLE}, we rewrite Eq. \eqref{eq:FWM1} to
\begin{equation}
\begin{aligned}
\frac{\partial}{\partial z}\left[\begin{matrix}\hat a_{as}\\\hat a_s^\dag\\\end{matrix}\right]
=\mathrm{M_{B}}\left[\begin{matrix}\hat a_{as}\\{\hat a_s^{\dag}}\\\end{matrix}\right]+\left[\begin{matrix}\hat F_{as}\\ -\hat F_s^{\dag}\\\end{matrix}\right],
\end{aligned}\label{eq:SFWM-BW}
\end{equation}
where
\begin{equation}
\begin{aligned}
&\mathrm{M_{B}}
=\left[\begin{matrix}-\alpha_{as}+i\frac{\Delta k}{2}&i\kappa\\i\kappa&\alpha_s^*-i\frac{\Delta k}{2}\\\end{matrix}\right],\\
&\hat F_{as}=\hat{\bar F}_{as}e^{-i\theta/2},\\
&\hat F_{s}^\dag=\hat{\bar F}_{s}^\dag e^{i\theta/2}.
\end{aligned}\label{eq:MBW1}
\end{equation}

Similarly, we rewrite the SFWM quantum Langevin equations in the forward-wave configuration in Appendix~\ref{Appendix: FW Quantum Langevin Equations}.

We now turn to compare Eq. \eqref{eq:SFWM-BW} with Eq. \eqref{eq:QLEBW} from the phenomenological approach in Sec. \ref{sec:QLE}, where we take mode 1 as anti-Stokes and mode 2 as Stokes in the backward-wave configuration. From Eq. \eqref{eq:QLEBW}, we have 
\begin{equation}
\begin{aligned}
\hat F_{as}&=
\mathrm{N_{BR}}_{11}\hat{f}_{1}
+\mathrm{N_{BI}}_{11}\hat{f}^{\dag}_{1}
+\mathrm{N_{BI}}_{12}\hat{f}_{2}
+\mathrm{N_{BR}}_{12}\hat{f}^{\dag}_{2},\\
\hat F_{s}^\dag&=-
\mathrm{N_{BR}}_{21}\hat{f}_{1}
-\mathrm{N_{BI}}_{21}\hat{f}^{\dag}_{1}
-\mathrm{N_{BI}}_{22}\hat{f}_{2}
-\mathrm{N_{BR}}_{22}\hat{f}^{\dag}_{2}.
\end{aligned}\label{eq:Fp}
\end{equation}
Therefore, we obtain $\hat F_{as}$ and $\hat F_{s}^\dag$ from two different approaches: Eq.~\eqref{eq:F} from the microscopic photon-atom interaction, and Eq.~\eqref{eq:Fp} from the macroscopic phenomenological approach. Although we remark that the atomic noise operators $\hat{\bar f}_{\mu\nu}^{(\sigma)}$ are different from the field noise operators $\hat f_m$, the correlations of $\hat F_{as}$ and $\hat F_{s}$ uniquely determine the system performance. While we find it difficult to analytically prove the two approaches are equivalent, we could numerically compute and compare the commutation relations and correlations of $\hat a_{as}$, $\hat a_{as}^{\dag}$, $\hat a_{s}$, and $\hat a_{s}^{\dag}$.  

We consider here the backward-wave SFWM in laser-cooled $^{85}$Rb atoms with relevant atomic energy levels being $\left|1\right>=\left|5^{2}\rm{S}_{1/2},\rm{F}=2\right>$, $\left|2\right>=\left|5^{2}\rm{S}_{1/2},\rm{F}=3\right>$, $\left|3\right>=\left|5^{2}\rm{P}_{1/2},\rm{F}=3\right>$, $\left|4\right>=\left|5^{2}\rm{P}_{3/2},\rm{F}=3\right>$. The decay and dephasing rates for corresponding energy levels are $\Gamma_3=\Gamma_4=2\pi\times6$ MHz, $\Gamma_{31}=\frac{5}{9}\Gamma_{3},\Gamma_{32}=\frac{4}{9}\Gamma_{3},\Gamma_{41}=\frac{4}{9}\Gamma_{4},\Gamma_{42}=\frac{5}{9}\Gamma_{4}$, $\gamma_{13}=\gamma_{23}=\gamma_{14}=\gamma_{24}=2\pi\times3$ MHz, and $\gamma_{12}=2\pi\times0.03$ MHz. With vacuum inputs in both Stokes ($z=L$) and anti-Stokes ($z=0$) modes, we have $\langle \hat a_{as}(\varpi,0)\hat a_{as}^{\dag}(\varpi',0)\rangle=\langle \hat a_{s}(\varpi,L)\hat a_{s}^{\dag}(\varpi',L)\rangle=\delta(\varpi-\varpi')$ and $\langle \hat a_{as}^{\dag}(\varpi,0)\hat a_{as}(\varpi',0)\rangle=\langle \hat a_{s}^{\dag}(\varpi,L)\hat a_{s}(\varpi',L)\rangle=0$. There is also no correlation between Stokes and anti-Stokes fields at their inputs.

We numerically compute SFWM in two different regimes to confirm the consistency between the macroscopic and microscopic theories. i) The first is the group delay regime, where the SFWM spectrum bandwidth is determined by the EIT slow-light induced phase mismatching \cite{du2008narrowband}. The working parameters are: $\Omega_p=2\pi\times1.2$ MHz, $\Omega_c=2\pi\times12$ MHz, $\Delta_p=2\pi\times500$ MHz. The cold atomic medium with length $L=2$ cm has density $n=5.1\times 10^{16}$ m$^{-3}$, corresponding to an atomic optical depth $\textrm{OD}=80$ on the anti-Stokes resonance transition. ii) The second is the Rabi oscillation regime, where biphoton correlation reveals single-atom dynamics \cite{du2008narrowband}. The working parameters are: $\Omega_p=2\pi\times1.2$ MHz, $\Omega_c=2\pi\times24$ MHz, $\Delta_p=\omega_p-\omega_{14}=2\pi\times500$ MHz. The cold atomic medium with length $L=0.2$ cm has density $n=6.4\times 10^{14}$ m$^{-3}$, corresponding to $\textrm{OD}=0.1$. In both cases, we take $\Delta k=127$ rad/m. 

The numerical results in the group delay regime are plotted in Figs.~\ref{fig:Fig03}, \ref{fig:Fig04}, and \ref{fig:Fig05}. The commutation relations $[\hat a_{as}(L), \hat a_{as}^{\dag}(L)]$ and $[\hat a_{s}(0), \hat a_{s}^{\dag}(0)]$ are shown in Fig.~\ref{fig:Fig03}. Both macroscopic and microscopic approaches agree well with each other [Figs. \ref{fig:Fig03}(a) and (c)], with negligible relative small difference $<1.0\times 10^{-6}$ [Figs.~\ref{fig:Fig03}(b) and (d)]. As expected, the macroscopic phenomenological results give perfect flat lines at $\frac{[\hat a_{as}(L, \varpi), \hat a^{\dag}_{as}(L, \varpi')]}{\delta(\varpi-\varpi')}=\frac{[\hat a_{s}(0, \varpi), \hat a^{\dag}_{s}(0, \varpi')]}{\delta(\varpi-\varpi')}=1$ which is the starting point of Sec. \ref{sec:QLE}. The microscopic results of field commutations are consistent with the macroscopic approach, but with $<1.0\times 10^{-6}$ deviation at some spectra points. As we understand, these small spectra discrepancies may be caused by the ground-state and zeroth-order approximations we take for solving the microscopic Heisenberg-Langevin equations \eqref{eq:HEM}. If the Langevin noise operators are not taken into account, as shown in the black dotted curves in Figs.~\ref{fig:Fig03}(a) and (c), the anti-Stokes commutation relation is not preserved and displays EIT transmission spectrum, while Stokes commutation relation still approximately holds due to the negligible gain or loss in Stokes channel under the ground-state approximation. 

Figure \ref{fig:Fig04} displays four real-valued correlations of Stokes and anti-Stokes fields: (a )$\langle \hat a_{as}(L)\hat a_{as}^{\dag}(L)\rangle$, (b) $\langle \hat a_{as}^{\dag}(L)\hat a_{as}(L)\rangle$, (c) $\langle \hat a_{s}(0)\hat a_{s}^{\dag}(0)\rangle$, and (d) $\langle \hat a_{s}^{\dag}(0)\hat a_{s}(0)\rangle$. Figure \ref{fig:Fig05} shows the twelve (six pairs) complex-valued correlations of Stokes and anti-Stokes fields: (a) $\langle \hat a_{as}(L)\hat a_{as}(L)\rangle=\langle \hat a_{as}^{\dag}(L)\hat a_{as}^{\dag}(L)\rangle^*$, (b) $\langle \hat a_{as}(L)\hat a_{s}(0)\rangle=\langle \hat a_{s}^{\dag}(0)\hat a_{as}^{\dag}(L)\rangle^*$, (c) $\langle \hat a_{as}(L)\hat a_{s}^{\dag}(0)\rangle=\langle \hat a_{s}(0)\hat a_{as}^{\dag}(L)\rangle^*$, (d) $\langle \hat a_{as}^{\dag}(L)\hat a_{s}(0)\rangle=\langle \hat a_{s}^{\dag}(0)\hat a_{as}(L)\rangle^*$, (e) $\langle \hat a_{s}(0)\hat a_{as}(L)\rangle=\langle \hat a_{as}^{\dag}(L)\hat a_{s}^{\dag}(0)\rangle^*$, and (f) $\langle \hat a_{s}(0)\hat a_{s}(0)\rangle=\langle \hat a_{s}^{\dag}(0)\hat a_{s}^{\dag}(0)\rangle^*$. The macroscopic solutions agree well with those obtained from the microscopic approach.

The numerical results in the Rabi oscillation regime are plotted in Figs.~\ref{fig:Fig06}, \ref{fig:Fig07}, and \ref{fig:Fig08}. The macroscopic phenomenological results also agree remarkably well with those from the microscopic theory.


\section{Biphoton Generation} \label{sec:Biphoton}

We now turn to apply the quantum Langevin theory to study time-frequency entangled photon pair (biphoton) generation through spontaneous four-wave mixing process, especially in a variety of situations involving gain, loss, and/or complex nonlinear coupling coefficient. We consider continuous-wave pumping whose time translation symmetry leads to frequency anti-correlation $\omega_1+\omega_2=$constant between the paired photons. In the spontaneous four-wave mixing process, both input states are vacuum: $\langle \hat a_1^\dag(\varpi,0)\hat a_1(\varpi',0)\rangle=\langle \hat a_2^\dag(\varpi,0)\hat a_2(\varpi',0)\rangle=0$, $\langle \hat a_1(\varpi',0)\hat a_1^\dag(\varpi,0)\rangle=\langle \hat a_2(\varpi',0)\hat a_2^\dag(\varpi,0)\rangle=\delta(\varpi-\varpi')$ for the forward-wave configuration, and $\langle \hat a_1^\dag(\varpi,0)\hat a_1(\varpi',0)\rangle=\langle \hat a_2^\dag(\varpi,L)\hat a_2(\varpi',L)\rangle=0$, $\langle \hat a_1(\varpi,0)\hat a_1^\dag(\varpi',0)\rangle=\langle \hat a_2(\varpi,L)\hat a_2^\dag(\varpi',L)\rangle=\delta(\varpi-\varpi')$ for the backward-wave configuration.  From Eq.~\eqref{eq:Fields}, with $\omega_{1}=\omega_{10}+\varpi$ and $\omega_2=\omega_{20}-\varpi$, we have
\begin{equation}
\begin{aligned}
&\hat a_{1}(t,z_1)
=\frac{e^{i\omega_{10}(\frac{z_1}{c}-t)}}{\sqrt{2\pi}}\int{d\varpi \hat a_{1}(\varpi,z_1)e^{i\varpi(\frac{z_1}{c}-t)}}e^{-i\frac{\Delta k}{2} z_1},\\
&\hat a_{2}(t,z_2)
=\frac{e^{i\omega_{20}(\pm\frac{z_2}{c}-t)}}{\sqrt{2\pi}}\int{d\varpi \hat a_{2}(\varpi,z_2)e^{i\varpi( \pm\frac{z_2}{c}-t)}}e^{-i\frac{\Delta k}{2} z_2},
\end{aligned}
\label{eq:Fields0}
\end{equation}
where $\pm$ represents the forward-wave ($+$) or backward-wave ($-$) configuration, $z=z_1$ and $z=z_2$ are the output positions of channels 1 and 2, respectively. For the forward-wave configuration, $z_1=z_2=L$. For the backward-wave configuration, $z_1=L$ and $z_2=0$. The phase mismatching in vacuum $\Delta k=(\omega_{as}\pm\omega_{s})/c-(\vec{k}_c+\vec{k}_p)\cdot \hat{z}\simeq (\omega_{as0}\pm\omega_{s0})/c-(\vec{k}_c+\vec{k}_p)\cdot \hat{z}$ is nearly a constant. The vacuum time delay $z_i/c$ constants are usually very small in usual experimental conditions, from now on we ignore these constants for simplification and rewrite the above equations to (otherwise one just needs to make a time translation $t\rightarrow t-z_i/c$)
\begin{equation}
\begin{aligned}
&\hat a_{1}(t,z_1)=\frac{e^{-i\omega_{10}t}}{\sqrt{2\pi}}\int{d\varpi \hat a_{1}(\varpi,z_1)e^{-i\varpi t}},\\
&\hat a_{2}(t,z_2)
=\frac{e^{-i\omega_{20}t}}{\sqrt{2\pi}}\int{d\varpi \hat a_{2}(\varpi,z_2)e^{i\varpi t}}.
\end{aligned}
\label{eq:Fields1}
\end{equation}
The photon rate in channel $m$ can be computed from
\begin{equation}
\begin{aligned}
R_{m}&\equiv\left<{\hat{a}}_{m}^\dag\left(t,z_m\right){\hat{a}}_{m}\left(t,z_m\right)\right>\\
&=\frac{1}{2\pi}\iint_{-\infty}^{\infty}{d\varpi d\varpi^\prime}e^{-i\varpi t}e^{i\varpi^\prime t}\left<{\hat{a}}_{m}^\dag\left(\varpi^\prime,z_m\right){\hat{a}}_{m}\left(\varpi,z_m\right)\right>.
\end{aligned}
\label{eq:Rate}
\end{equation}
Here we are particularly interested in the two-photon Glauber correlation in the time domain, which can be computed from the following two different orders
\begin{equation}
\begin{aligned}
&G_{2,1}^{\left(2\right)}\left(t_{2},t_{1}\right)\\
\equiv&\langle\hat{a}_{1}^\dag\left(t_{1},z_1\right){\hat{a}}_{2}^\dag\left(t_2,z_2\right){\hat{a}}_{2}\left(t_{2},z_2\right){\hat{a}}_{1}\left(t_{1},z_1\right)\rangle\\
=&|\langle\hat{a}_{2}\left(t_2,z_2\right)\hat{a}_{1}\left(t_{1},z_1\right)\rangle|^2\\
&+|\langle\hat{a}_{2}^\dag\left(t_{2},z_2\right){\hat{a}}_{1}\left(t_{1},z_1\right)\rangle|^2
+R_{1}R_{2},
\end{aligned}
\label{eq:G2-1}
\end{equation}
\begin{equation}
\begin{aligned}
&G_{1,2}^{\left(2\right)}\left(t_{1},t_{2}\right)\\
\equiv&\langle{\hat{a}}_{2}^\dag\left(t_2,z_2\right)\hat{a}_{1}^\dag\left(t_{1},z_1\right){\hat{a}}_{1}\left(t_{1},z_1\right){\hat{a}}_{2}\left(t_{2},z_2\right)\rangle\\
=&|\langle\hat{a}_{1}\left(t_1,z_1\right)\hat{a}_{2}\left(t_{2},z_2\right)\rangle|^2\\
&+|\langle\hat{a}_{2}^\dag\left(t_{2},z_2\right){\hat{a}}_{1}\left(t_{1},z_1\right)\rangle|^2
+R_{1}R_{2},
\end{aligned}
\label{eq:G1-2}
\end{equation}
where we have applied the Gaussian moment theorem \cite{Mandel, Louisell} to decompose the fourth-order field correlations to the sum of the products of second-order field correlations. The first term in Eqs. \eqref{eq:G2-1} and \eqref{eq:G1-2} can be expressed as $|\Psi_{2,1}(t_2,t_1)|^2$ and $|\Psi_{1,2}(t_1,t_2)|^2$, where 
\begin{equation}
\begin{aligned}
\Psi_{2,1}(t_2,t_1)&=\left<{\hat{a}}_{2}\left(t_2,z_2\right){\hat{a}}_{1}\left(t_{1},z_1\right)\right>\\
&=e^{-i\omega_{20}t_2}e^{-i\omega_{10}t_1}\psi_{2,1}(t_1-t_2),
\end{aligned}
\label{eq:Psi2-1}
\end{equation}
\begin{equation}
\begin{aligned}
\Psi_{1,2}(t_1,t_2)&=\left<{\hat{a}}_{1}\left(t_1,z_1\right){\hat{a}}_{2}\left(t_{2},z_2\right)\right>\\
&=e^{-i\omega_{20}t_2}e^{-i\omega_{10}t_1}\psi_{1,2}(t_1-t_2),
\end{aligned}
\label{eq:Psi1-2}
\end{equation}
are the two-photon wavefunctions with the relative parts
\begin{equation}
\begin{aligned}
&\psi_{2,1}(t_1-t_2)\\
&=\frac{1}{2\pi}\iint d\varpi d\varpi' \langle \hat a_2(\varpi',z_2)\hat a_1(\varpi, z_1)\rangle e^{-i\varpi (t_1-t_2)}.
\end{aligned}
\label{eq:psi2-1}
\end{equation}
\begin{equation}
\begin{aligned}
&\psi_{1,2}(t_1-t_2)\\
&=\frac{1}{2\pi}\iint d\varpi d\varpi' \langle \hat a_1(\varpi,z_1)\hat a_2(\varpi', z_2)\rangle e^{-i\varpi (t_1-t_2)}.
\end{aligned}
\label{eq:psi1-2}
\end{equation}
One can show that the second term in Eqs.~\eqref{eq:G2-1} and \eqref{eq:G1-2} is zero if the nonlinear coupling coefficient is real-valued, and it is usually very small as compared to other terms. The third term in Eqs.~\eqref{eq:G2-1} and \eqref{eq:G1-2} is the accidental coincidence counts. The two-photon wavefunction and Glauber correlation satisfy the following exchange symmetry
\begin{equation}
\begin{aligned}
\psi_{21}(t_1-t_2)=\psi_{2,1}(t_1-t_2)=\psi_{1,2}(t_1-t_2), \\
\Psi_{21}(t_2,t_1)=\Psi_{2,1}(t_2, t_1)=\Psi_{1,2}(t_1, t_2), \\
G_{21}^{\left(2\right)}\left(t_{2},t_{1}\right)=G_{2,1}^{\left(2\right)}\left(t_{2},t_{1}\right)=G_{1,2}^{\left(2\right)}\left(t_{1},t_{2}\right).
\end{aligned}
\label{eq:G2symmetry}
\end{equation}
The normalized two-photon correlation is defined as 
\begin{equation}
\begin{aligned}
g_{21}^{\left(2\right)}\left(t_{2},t_{1}\right)\equiv\frac{G_{21}^{\left(2\right)}\left(t_{2},t_{1}\right)}{R_1 R_2}.
\end{aligned}
\label{eq:g2}
\end{equation}
As the system has time translation symmetry with continuous-wave pumping, $G_{21}^{\left(2\right)}\left(t_{2},t_{1}\right)=G_{21}^{\left(2\right)}\left(t_{1}-t_{2}\right)$ depends only on the relative time $t_1-t_2$. 

\subsection{Loss and Gain} \label{sec:TimeOrder}

To simplify and unify the descriptions for accounting both forward- and backward-wave cases, we define ``input-output'' fields: $\hat a_{1,in}\equiv\hat a_1(0)$, $\hat a_{2,in}\equiv\hat a_2(0)$, $\hat a_{1,out}\equiv\hat a_1(L)$, and $\hat a_{2,out}\equiv\hat a_2(L)$ for the forward-wave case; $\hat a_{1,in}\equiv\hat a_1(0)$, $\hat a_{2,in}\equiv\hat a_2(L)$, $\hat a_{1,out}\equiv\hat a_1(L)$, and $\hat a_{2,out}\equiv\hat a_2(0)$ for the backward-wave case. In this subsection, we aim to investigate the roles of loss and gain in biphoton generation, considering linear loss in mode 1 ($\mathrm{Re}\{\alpha_1\}=\alpha\geq 0$) and linear gain ($\mathrm{Re}\{\alpha_2\}=-g\leq 0$) in mode 2. We also assume $\kappa$ is real, or its contribution to Langevin noises is much smaller than the linear gain and loss, \textit{i.e.},  $\mathrm{Im}\{\kappa\} \ll \{\alpha, g\}$. In this case, for forward- and backward-wave configurations, the noise matrix is reduced to
\begin{equation}
\begin{aligned}
\mathrm{N_{F,B}}=\left[\begin{matrix}\sqrt{2\alpha}&0\\0&\pm i\sqrt{2 g}\\\end{matrix}\right].
\end{aligned}\label{eq:NFBW}
\end{equation}
Hence, the output fields in Eqs. \eqref{eq:SolutionFW2} and \eqref{eq:SolutionBW2} can be rewritten as 
\begin{equation}
\begin{aligned}
\left[\begin{matrix}\hat a_{1,out}\\\hat a_{2,out}^\dag\\\end{matrix}\right]
=\left[\begin{matrix}A&B\\C&D\\\end{matrix}\right]\left[\begin{matrix}\hat a_{1,in}\\\hat a_{2,in}^\dag\\\end{matrix}\right]+\int_{0}^{L}\left[\begin{matrix}X_{11}&X_{12}\\X_{21}&X_{22}\\\end{matrix}\right]\left[\begin{matrix}\hat f_{1}\left(z\right)\\\hat f_2\left(z\right)\\\end{matrix}\right]dz.
\end{aligned}\label{eq:Solutionlg}
\end{equation}
where $X_{mn}$ are combined coefficients. We further rewrite Eq. \eqref{eq:Solutionlg} as
\begin{equation}
    \begin{aligned}
        &\hat a_{1,out}=A\hat a_{1,in}+B\hat a_{2,in}^\dag +\int_0^L\left[X_{11}\hat f_1(z)+X_{12}\hat f_2(z)\right],\\
        &\hat a_{2,out}=C^*\hat a_{1,in}^\dag+D^*\hat a_{2,in} +\int_0^L\left[X_{21}^*\hat f_1^\dag(z)+X_{22}^*\hat f_2^\dag(z)\right].
    \end{aligned}
    \label{eq:Solutionlg2}
\end{equation}
As shown in Eq. \eqref{eq:G2symmetry}, there are two different orders [$\langle : \hat a_2 \hat a_1 :\rangle$ or $\langle : \hat a_1 \hat a_2 :\rangle$] to compute the two-photon wavefunction and Galuber correlation. Although these two orders are equivalent, the numerical computation complexity may be significantly different. Computing biphoton wavefunction in Eq. \eqref{eq:psi1-2} in the order $\langle : \hat a_1 \hat a_2 :\rangle$ involves nonzero noise field correlations $\langle\hat f_m \hat f_m^\dag\rangle$, while in the order $\langle : \hat a_2 \hat a_1 :\rangle$ [Eq. \eqref{eq:psi2-1}] these noise field correlations disappear because of $\langle\hat f_m^\dag \hat f_m\rangle=0$. These field correlations in the frequency domain can be expressed as
\begin{equation}
\begin{aligned}
&\left<{\hat{a}}_{2out}\left(\varpi^\prime\right){\hat{a}}_{1out}\left(\varpi\right)\right>
=\delta(\varpi-\varpi^\prime)\left[{{B}}{{D}^\ast}\right],
\end{aligned}\label{eq:Csas}
\end{equation}
\begin{equation}
\begin{aligned}
&\left<{\hat{a}}_{1out}\left(\varpi\right){\hat{a}}_{2out}\left(\varpi^\prime\right)\right>\\
&=\delta(\varpi-\varpi^\prime)\left[{{A}}{{C}^\ast}+\int_{0}^{L} dz\left( X_{11}X_{21}^*+X_{12}X_{22}^*\right)\right].
\end{aligned}\label{Cass}
\end{equation}
Therefore, we obtain the biphoton wavefunction following the order $\langle : \hat a_2 \hat a_1 :\rangle$
\begin{equation}
\begin{aligned}
\psi_{21}(\tau)&=\iint d\varpi d\varpi' \langle \hat a_{2,out}(\varpi')\hat a_{1,out}(\varpi)\rangle e^{-i\varpi \tau}\\
&=\int d\varpi BD^*e^{-i\varpi \tau}.
\end{aligned}
\label{eq:psi2-1BD}
\end{equation}
where $\tau=t_1-t_2$. If following the order $\langle : \hat a_1 \hat a_2 :\rangle$, we have 
\begin{equation}
\begin{aligned}
\psi_{12}(\tau)&=\iint d\varpi d\varpi' \langle \hat a_{1,out}(\varpi)\hat a_{2,out}(\varpi')\rangle e^{-i\varpi \tau}\\
&=\int d\varpi \left[{{A}}{{C}^\ast}+\int_{0}^{L} dz\left( X_{11}X_{21}^*+X_{12}X_{22}^*\right)\right]e^{-i\varpi\tau}.
\end{aligned}
\label{eq:psi1-2AC}
\end{equation}
One can show that the second term in Eqs. \eqref{eq:G2-1} and \eqref{eq:G1-2} is zero in this loss-gain configuration. The single-channel photon rates can be obtained as
\begin{equation}
    \begin{aligned}
        R_1&=\frac{1}{2\pi}\int |B|^2 d\varpi,\\
        R_2&=\frac{1}{2\pi}\int \left[|C|^2+\int_{0}^{L} dz\left( |X_{21}|^2+|X_{22}|^2\right)\right] d\varpi.
    \end{aligned}
\end{equation}

It is interesting to remark that, in the loss-gain configuration, the biphoton field correlation following the order $\langle:\hat a_{\rm{gain}} \hat a_{\rm{loss}}:\rangle$ does not involve noise field correlations as shown in Eqs. \eqref{eq:Csas} and \eqref{eq:psi2-1BD}, which dramatically reduces the computation complexity. On the other side, taking the order $\langle:\hat a_{\rm{loss}} \hat a_{\rm{gain}}:\rangle$ must include noise field correlations as shown in Eqs. \eqref{Cass} and \eqref{eq:psi1-2AC}. This may be understood in the heralded photon picture \cite{PhysRevA.92.043836}: When a photon in a lossy channel is detected (annihilated) by a detector, we can always ensure there is its partner (or paired) photon in another channel; On the other side, when a photon is detected in a gain channel which produces multiple photons, we can not always ensure it has a partner photon in another channel. The exchange symmetry can only be preserved by taking into account the Langevin noises. 

\begin{figure}
\centering
\includegraphics{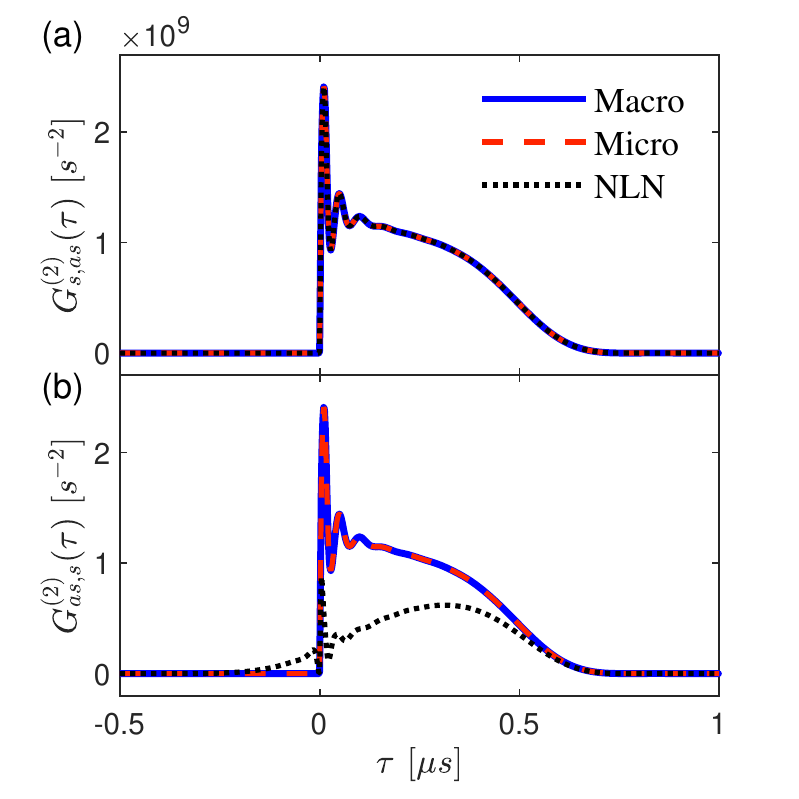}
\caption{Two-photon Glauber correlation in time domain in the group delay regime: (a) $G_{s,as}^{(2)}(\tau)$ and (b) $G_{as,s}^{(2)}(\tau)$. The simulation conditions are the same as that in Figs.~\ref{fig:Fig03}, \ref{fig:Fig04}, and \ref{fig:Fig05}. NLN: no Langevin noise included.}
\label{fig:Fig09}
\end{figure}

In the SFWM described in Sec. \ref{sec:Microscopic}, the anti-Stokes photons experience finite EIT loss due to the ground state dephasing ($\gamma_{12}\neq 0$), and the Stokes photons propagate with negligible but small Raman gain. Figure \ref{fig:Fig09} displays the two-photon Glauber correlation in the group delay regime with the same parameters as those in Figs. \ref{fig:Fig03}, \ref{fig:Fig04} and \ref{fig:Fig05}. As shown in Fig.~\ref{fig:Fig09}(a) and (b), both macroscopic and microscopic approaches with Langevin noises give consistent results. As expected, the computation of $G_{s,as}^{(2)}(\tau)$ (following the order $\langle:\hat a_{s} \hat a_{as}:\rangle$) without Langevin noise operators (black dotted line: NLN) agrees with the exact results obtained from both macroscopic (blue solid line) and microscopic (red dashed line) approaches, shown in Fig.~\ref{fig:Fig09}(a). On the contrary, the computation of $G_{as,s}^{(2)}(\tau)$ (following the order $\langle:\hat a_{as} \hat a_{s}:\rangle$) without Langevin noise operators deviates significantly from the exact results, as shown in Fig.~\ref{fig:Fig09}(b). 

\subsection{Complex Phase Mismatching}\label{Complex Phase Mismatching}

Different from the Heisenberg picture where the evolution of field operators is governed by their Langevin coupled equations, reference \cite{du2008narrowband} provides a perturbation theory to describe biphoton state in the interaction picture. The solution from Heisenberg-Langevin theory may contain correlations of more than two photons, while the perturbation theory focuses only on the two-photon state by ignoring higher-order terms. These two treatments are expected to give the same results in the limit of small parameter gain. Although the perturbation theory in the interaction picture provides a much clear physics picture of two-photon state, treating loss and gain requires a proper justification. In the perturbation theory, linear loss and gain are included in the complex phase mismatching $\Delta\tilde k(\varpi)$ \cite{du2008narrowband}. For the SFWM described in Sec. \ref{sec:Microscopic}, Ref. \cite{du2008narrowband} derives the biphoton relative wavefunction with perturbation theory as
\begin{equation}
    \begin{aligned}
        \psi(\tau)=\frac{iL}{2\pi}\int d\varpi \kappa(\varpi)\Phi(\varpi)e^{-i\varpi\tau},
    \end{aligned}
    \label{eq:psiJOSAB}
\end{equation}
where the longitudinal detuning function is
\begin{equation}
    \begin{aligned}
        \Phi(\varpi)=\mathrm{sinc}\left(\frac{\Delta\tilde k L}{2}\right)e^{i(k_{as}+k_{s})L},
    \end{aligned}
    \label{eq:PhiJOSAB}
\end{equation}
There is a statement in Ref.~\cite{du2008narrowband}: ``It is found that to be consistent with the Heisenberg–Langevin theory in the low-gain limit, the argument in $\Phi$ should be replaced by $\Delta\tilde k = \left(\vec{k}_{as}+\vec{k}_s^\ast-\vec{k}_c-\vec{k}_p \right)\cdot \hat{z}$, where $\vec{k}_s^\ast$ is the conjugate of $\vec{k}_s$." For the SFWM in the double-$\Lambda$ four-level atomic system, there is small Raman gain in the Stokes channel. What happens if there is loss in the Stokes channel? Should we take $\vec{k}_s^\ast$ or $\vec{k}_s$ in the complex phase mismatching $\Delta\tilde k(\varpi)$? Although Ref.~\cite{du2008narrowband} takes $\vec{k}_s^\ast$ for Stokes photons with gain, it is not clear whether it still holds for the case with loss. In this subsection, we do not only provide a justification for the above statement in Ref.~\cite{du2008narrowband} from the quantum Langevin theory by taking small parametric gain approximation, but also extend the complex phase mismatching to the case with loss in the Stokes channel.     

We take the same backward-wave configuration in Ref.~\cite{du2008narrowband}. We assume anti-Stokes photons in mode 1 are lossless with EIT and there is gain (or loss) in Stokes mode 2. The small parametric gain fulfills $\left|\kappa\right| \ll \{\alpha, g\}$.

In the backward-wave configuration, using Eq. (\ref{eq:MBW}), (\ref{ABCDBW}), and (\ref{barABCD}), we obtain analytical expressions of $A, B, C$, and $D$ as
\begin{equation}
\begin{aligned}
A &= \frac{\sqrt{q^2-4\kappa^2}e^{-(\alpha_1-\alpha_2^*)L/2}}{q\mathrm{sinh}\left(\frac{L}{2}\sqrt{q^2-4\kappa^2}\right)+\sqrt{q^2-4\kappa^2}\mathrm{cosh}\left(\frac{L}{2}\sqrt{q^2-4\kappa^2}\right)},\\
B &= \frac{2i\kappa}{q+\sqrt{q^2-4\kappa^2}\mathrm{coth}(\frac{L}{2}\sqrt{q^2-4\kappa^2})},\\
C &= \frac{-2i\kappa}{q+\sqrt{q^2-4\kappa^2}\mathrm{coth}(\frac{L}{2}\sqrt{q^2-4\kappa^2})},\\
D &= \frac{\sqrt{q^2-4\kappa^2}e^{(\alpha_1-\alpha_2^*)L/2}}{q\mathrm{sinh}\left(\frac{L}{2}\sqrt{q^2-4\kappa^2}\right)+\sqrt{q^2-4\kappa^2}\mathrm{cosh}\left(\frac{L}{2}\sqrt{q^2-4\kappa^2}\right)},
\end{aligned}\label{ABCDexpr}
\end{equation}
where $q\equiv\alpha_1+\alpha_2^*-i\Delta k$. In the small parametric gain approximation, we have
\begin{equation}
\begin{aligned}
&\sqrt{q^2-4\kappa^2}\approx q\\ &=\alpha_1+\alpha_2^*-i\Delta k
=-i(\Delta{k_1}-\Delta{k_2}^*+\Delta k),
\end{aligned}
\end{equation}
and
\begin{equation}
\begin{aligned}
\alpha_1-\alpha_2^*
=-i(\Delta{k_1}+\Delta{k_2}^*).
\end{aligned}
\end{equation}
where $\Delta k_m=\frac{\omega_m}{2c}\chi_{m}$ is the wavenumber difference from that in vacuum. Hence, we simplify $A, B, C$, and $D$ to
\begin{equation}
\begin{aligned}
A =& \mathrm{exp}\left[i\Delta k_1 L\right]\mathrm{exp}\left[\frac{i\Delta k L}{2}\right],\\
B =&i\kappa L \mathrm{sinc}\left[\frac{(\Delta k_1-\Delta k_2^*+\Delta k)L}{2}\right]\\
&\times\mathrm{exp}\left[\frac{i(\Delta k_1-\Delta k_2^*+\Delta k)L}{2}\right],\\
C =&-i\kappa L \mathrm{sinc}\left[\frac{(\Delta k_1-\Delta k_2^*+\Delta k)L}{2}\right]\\
&\times\mathrm{exp}\left[\frac{i(\Delta k_1-\Delta k_2^*+\Delta k)L}{2}\right],\\
D =&\mathrm{exp}\left[-i\Delta k_2^* L\right]\mathrm{exp}\left[\frac{i\Delta k L}{2}\right].
\end{aligned}\label{ABCDexpr2}
\end{equation}

We first look at the case with gain in the Stokes (mode 2). As discussed in Sec. \ref{sec:TimeOrder}, we take the order $\langle:\hat a_{2}\hat a_{1}:\rangle$ 
\begin{equation}
\begin{aligned}
\psi_{21}(\tau)&=\iint d\varpi d\varpi' \langle \hat a_{2,out}(\varpi')\hat a_{1,out}(\varpi)\rangle e^{-i\varpi \tau}\\
&=\int d\varpi BD^*e^{-i\varpi \tau},
\end{aligned}
\label{eq:psigain1}
\end{equation}
where
\begin{equation}
\begin{aligned}
{{B}}{{D}^\ast}
&=i\kappa L\mathrm{sinc}\left[\frac{(\Delta k_1-\Delta k_2^*+\Delta k)L}{2}\right]\\
&\times\mathrm{exp}\left[\frac{i(\Delta k_1-\Delta k_2^*+2\Delta k_2)L}{2}\right].
\end{aligned}\label{eq:BDgain}
\end{equation}
Comparing Eqs.~\eqref{eq:psigain1} and \eqref{eq:BDgain} with Eqs.~\eqref{eq:psiJOSAB} and \eqref{eq:PhiJOSAB}, particularly for the argument in the $\mathrm{sinc}$ function, we have $\Delta\tilde k=\Delta k_1-\Delta k_2^*+\Delta k=k_{1}-k_2^*-k_c+k_p=k_{as}-k_s^*-k_c+k_p$ which is consistent with the statement in Ref.~\cite{du2008narrowband}. 

We now look at the case with loss in the Stokes (mode 2). We take the order $\langle:\hat a_{1}\hat a_{2}:\rangle$ and have 
\begin{equation}
\begin{aligned}
\psi_{12}(\tau)&=\iint d\varpi d\varpi' \langle \hat a_{1,out}(\varpi)\hat a_{2,out}(\varpi')\rangle e^{-i\varpi \tau}\\
&=\int d\varpi AC^*e^{-i\varpi \tau},
\end{aligned}
\label{eq:psiloss1}
\end{equation}
where
\begin{equation}
\begin{aligned}
{{A}}{{C}^\ast}&=i\kappa^\ast L\mathrm{sinc}\left[\frac{(\Delta k_1^*-\Delta k_2+\Delta k)L}{2}\right]\\
&\mathrm{exp}\left[\frac{i(2\Delta k_1-\Delta k_1^*+\Delta k_2)L}{2}\right].
\end{aligned}\label{eq:ACloss}
\end{equation}
Comparing Eqs.~\eqref{eq:psiloss1} and \eqref{eq:ACloss} with Eqs.~\eqref{eq:psiJOSAB} and \eqref{eq:PhiJOSAB}, we have $\Delta\tilde k=\Delta k_1^*-\Delta k_2+\Delta k=k_{1}-k_2-k_c+k_p=k_{as}-k_s-k_c+k_p$, which is different from the case with gain. Here we have taken $k_1\simeq k_1^*$ for lossless mode 1.

Although our discussion is based on the backward-wave configuration, the conclusion can be extended to the forward-wave configuration, which is derived in detail in Appendix \ref{Appendix: FW Complex Phase Mismatching}. Therefore, in the case with gain in the Stokes mode 2, the complex phase mismatching is $\Delta\tilde k = \left(\vec{k}_{as}+\vec{k}_s^\ast-\vec{k}_c-\vec{k}_p \right)\cdot \hat{z}$. In the case with loss in the Stokes mode 2, the complex phase mismatching becomes $\Delta\tilde k = \left(\vec{k}_{as}+\vec{k}_s-\vec{k}_c-\vec{k}_p \right)\cdot \hat{z}$.


\subsection{Complex Nonlinear Coupling Coefficient and Rabi Oscillation}

As illustrated in Fig.~\ref{fig:Fig02}, we can understand the SFWM process in the following picture. After a Stoke and anti-Stokes photon pair is born from a single atom following the atomic transitions [Fig.~\ref{fig:Fig02}(b)], the paired photons then propagate through the medium [Fig.~\ref{fig:Fig02}(a)]. As the photon pair can be generated at any atom inside the medium, the overall two-photon wavefunction (or probability amplitude) is a superposition of all possible such generation-propagation two-photon Feynman paths. Following this picture, when the propagation effect can be ignored, the biphoton state reveals the single atom dynamics, which is connected to the nonlinear coupling coefficient. In the following, we consider SFWM in the limit of small optical depth (OD) where the linear propagation effect is small and show how the complex spectrum of nonlinear coupling coefficient reveals single-atom Rabi oscillation.

We rewrite the nonlinear coupling coefficient in Eq.~\eqref{eq:kappa} as:
\begin{equation}
\begin{aligned}
\kappa(\varpi)=J\left[\frac{1}{(\varpi-\Omega_e/2+i\gamma_e)}-\frac{1}{ (\varpi+\Omega_e/2+i\gamma_e)}\right],
\end{aligned}\label{eq:kappa4}
\end{equation}
where
\begin{equation}
\begin{aligned}
J=-\frac{\sqrt{\omega_{as}\omega_s}n\mu_{13}\mu_{24}}{8c\varepsilon_0\hbar\Omega_e}\left|\frac{\Omega_p \Omega_c}{\Delta_p+i\gamma_{14}}\right|.
\end{aligned}\label{eq:J}
\end{equation}
Here $\Omega_e=\sqrt{|\Omega_c|^2-(\gamma_{13}-\gamma_{12})^2}$ is the effective coupling Rabi frequency, and $\gamma_e=(\gamma_{12}+\gamma_{13})/2$ is the effective dephasing rate. Obviously, the nonlinear coupling coefficient $\kappa(\varpi)$ has a complex spectrum, with two resonances separated by the effective coupling Rabi frequency $\Omega_e$. In the ground-state approximation with major atomic population in state $|1\rangle$, the undepleted pump laser beam is far detuned from the transition $|1\rangle\rightarrow|4\rangle$ and its excitation is weak such that we can take $\chi_s\simeq 0$. On the other side, from Eq. \eqref{eq:parameters} we have the complex linear susceptibility for anti-Stokes photons   
\begin{equation}
\begin{aligned}
\chi_{as}(\varpi)=-\frac{n\left|\mu_{13}\right|^2}{\varepsilon_0\hbar}\frac{\left(\varpi+i\gamma_{12}\right)}{(\varpi-\Omega_e/2+i\gamma_e)(\varpi+\Omega_e/2+i\gamma_e)}
\end{aligned}\label{eq:chialinear}
\end{equation}
Although the anti-Stokes photon absorption at $\varpi=0$ is suppressed by the EIT effect, there are two absorption resonances appearing at $\varpi=\pm\Omega_e/2$ which coincide with the two resonances of nonlinear coupling coefficient in Eq.~\eqref{eq:kappa4}. We take the pump laser with weak intensity ($\propto |\Omega_p|^2$) and large detuning ($\Delta_p$) such that Re\{$\alpha_{as}(\varpi=\pm\Omega_e/2)$\}$>$Im\{$\kappa(\varpi=\pm\Omega_e/2)$\}, which are usually satisfied in the ground state condition. As the propagation effect is small and the phase matching is not important, the paired photons are mostly generated from the two resonances ($\varpi=\pm\Omega_e/2$) of the nonlinear coupling coefficient. 

In the forward-wave configuration, with the coupling matrix 
\begin{equation}
\begin{aligned}
\mathrm{M_F}
=\left[\begin{matrix}-\alpha_{as}+i\frac{\Delta k}{2}&i\kappa\\-i\kappa&-i\frac{\Delta k}{2}\\\end{matrix}\right],
\end{aligned}\label{eq:MFRabi}
\end{equation}
and short medium length $L$ satisfying $|\mathrm{M_F}L|\ll 1$, we have approximately 
\begin{equation}
\begin{aligned}
\left[\begin{matrix}A&B\\C&D\\\end{matrix}\right]&=e^{\mathrm{M_{F}}L}\cong \mathbb{1}+\mathrm{M_{F}}L\\
&=\left[\begin{matrix}1-\alpha_{as}L+i\frac{\Delta k}{2}L&i\kappa L\\-i\kappa L&1-i\frac{\Delta k}{2}L\\\end{matrix}\right].
\end{aligned}\label{ABCDFWRabi}
\end{equation}
As discussed in Sec. \ref{sec:TimeOrder}, the biphoton field correlation following the order $\langle:\hat a_{s}\hat a_{as}:\rangle$ does not need count the Langevin noise operators:
\begin{equation}
\begin{aligned}
\langle \hat a_{s}(\varpi',L)\hat a_{as}(\varpi,L)\rangle
&=BD^*\delta(\varpi-\varpi')\\
&=i\kappa L(1+i\frac{\Delta k}{2}L)\delta(\varpi-\varpi')\\
&\cong i\kappa(\varpi) L \delta(\varpi-\varpi'),
\end{aligned}\label{eq:BiphotonWavefunctionFrequencyFW}
\end{equation}
where we have neglected higher order terms $\mathrm{O}(L^2)$. From Eq.~\eqref{eq:psi2-1}, 
we have the relative biphoton wavefunction
\begin{equation}
\begin{aligned}
\psi_{s-as}(\tau)=\frac{iL}{2\pi}\int d\varpi \kappa(\varpi) e^{-i\varpi\tau},
\end{aligned}
\label{eq:BiphtonRelativeWavefuncitonRabiFW00}
\end{equation}
which is the Fourier transform of the nonlinear coupling coefficient with $\tau=t_{as}-t_s$. Substituting Eq. \eqref{eq:kappa4} into Eq. \eqref{eq:BiphtonRelativeWavefuncitonRabiFW00} we obtain
\begin{equation}
\begin{aligned}
\psi_{s-as}(\tau)&=L J e^{-\gamma_e \tau}[e^{-i\Omega_e\tau/2}-e^{i\Omega_e\tau/2}]\Theta(\tau) \\
&=-2iL J e^{-\gamma_e \tau} \sin{\left(\frac{\Omega_e\tau}{2}\right)}\Theta(\tau),
\end{aligned}
\label{eq:BiphtonRelativeWavefuncitonRabiFW01}
\end{equation}
where $\Theta(\tau)$ is the Heaviside function. Equation \eqref{eq:BiphtonRelativeWavefuncitonRabiFW01} shows a damped Rabi oscillation, resulting from the beating between biphotons generated from the two resonances at $\varpi=\pm\Omega_e/2$. The Heaviside function shows the anti-Stokes photon is always generated after its paired Stokes photon following the time order of atomic transitions $|1\rangle\rightarrow|4\rangle\rightarrow|2\rangle\rightarrow|3\rangle\rightarrow|1\rangle$ in an SFWM cycle shown in Fig. \ref{fig:Fig02}(b). 

\begin{figure}
\centering
\includegraphics{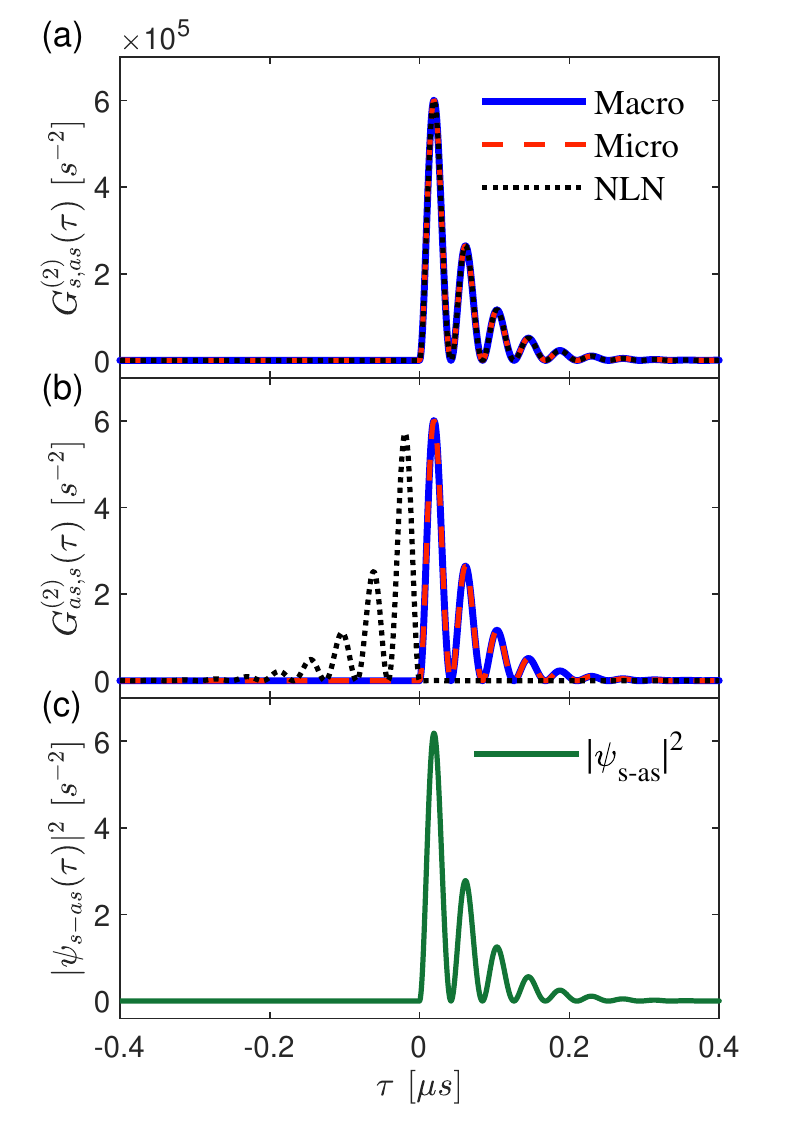}
\caption{Two-photon Glauber correlation in time domain in the damped Rabi oscillation regime: (a) $G_{s,as}^{(2)}(\tau)$ and (b) $G_{as,s}^{(2)}(\tau)$. The simulation conditions are the same as that in Figs.~\ref{fig:Fig06}, \ref{fig:Fig07}, and \ref{fig:Fig08}. (c) shows the analytic solution for the biphoton waveform $\left|\psi_{s-as}(\tau)\right|^2$. NLN: no Langevin noise included.}
\label{fig:Fig10}
\end{figure}

In the backward-wave configuration, the coupling matrix becomes
\begin{equation}
\begin{aligned}
\mathrm{M_B}
=\left[\begin{matrix}-\alpha_{as}+i\frac{\Delta k}{2}&i\kappa\\i\kappa&-i\frac{\Delta k}{2}\\\end{matrix}\right].
\end{aligned}\label{eq:MBRabi}
\end{equation}
With $|\mathrm{M_B}L|\ll 1$ we have 
\begin{equation}
\begin{aligned}
\left[\begin{matrix}\bar A&\bar B\\\bar C&\bar D\\\end{matrix}\right]&=e^{\mathrm{M_{B}}L}\cong \mathbb{1}+\mathrm{M_{B}}L\\
&=\left[\begin{matrix}1-\alpha_{as}L+i\frac{\Delta k}{2}L&i\kappa L\\i\kappa L&1-i\frac{\Delta k}{2}L\\\end{matrix}\right],
\end{aligned}\label{ABCDBWRabi}
\end{equation}
and 
\begin{equation}
\begin{aligned}
\left[\begin{matrix}A&B\\C&D\\\end{matrix}\right]=\left[\begin{matrix}1-\alpha_{as}L+i\frac{\Delta k}{2}L&i\kappa L\\-i\kappa L&1+i\frac{\Delta k}{2}L\\\end{matrix}\right],
\end{aligned}\label{eq:A1B1C1D1BWRabi}
\end{equation}
where we have neglect higher order terms $\mathrm{O}(L^2)$. Similarly, we have
\begin{equation}
\begin{aligned}
\langle \hat a_{s}(\varpi',0)\hat a_{as}(\varpi,L)\rangle\cong i\kappa(\varpi) L \delta(\varpi-\varpi'),
\end{aligned}\label{eq:BiphotonWavefunctionFrequencyBW}
\end{equation}
which is the same as Eq. \eqref{eq:BiphotonWavefunctionFrequencyFW} of the forward-wave configuration. Therefore, we obtain Rabi oscillations in both forward- and backward-wave configurations. Equation \eqref{eq:BiphtonRelativeWavefuncitonRabiFW01} is identical to the result derived from the perturbation theory in the interaction picture \cite{du2008narrowband}.  

Figure \ref{fig:Fig10} displays the two-photon Glauber correlation in the damped Rabi oscillation regime with the same parameters as those in Figs. \ref{fig:Fig06}, \ref{fig:Fig07} and \ref{fig:Fig08}. As illustrated in Fig.~\ref{fig:Fig10}(a) and (b), both macroscopic and microscopic approaches with Langevin noises give consistent results. As expected, the computation of $G_{s,as}^{(2)}(\tau)$ (following the order $\langle:\hat a_{s} \hat a_{as}:\rangle$) without Langevin noise operators (dot points) agrees with the exact results obtained from both microscopic (red dashed line) and macroscopic (blue solid line) approaches, shown in Fig.~\ref{fig:Fig10}(a). On the contrary, the computation of $G_{as,s}^{(2)}(\tau)$ (following the order $\langle:\hat a_{as} \hat a_{s}:\rangle$) without Langevin noise operators (dot points: NLN) deviates significantly from the exact results and violates the causality, as shown in Fig.~\ref{fig:Fig10}(b). Fig.~\ref{fig:Fig10}(c) shows the result from the analytic solution in Eq.~\eqref{eq:BiphtonRelativeWavefuncitonRabiFW01} which agree well with the exact results in Figs.~\ref{fig:Fig10}(a) and (b).

It is interesting to examine a system without gain and loss whose Langevin noises are purely contributed by the complex nonlinear coupling coefficient. In this case, the above approximation and conclusion do not hold. Let's now consider the case 3 with the forward-wave configuration in Sec.~\ref{FW system}, where $\alpha_1=\alpha_2=\Delta k =0$, and $\kappa=\eta+i\zeta$. As shown in Sec.~\ref{FW system}, the noise matrix is different as $\zeta$ is positive or negative. We first consider $\zeta >0$, the Langevin coupled equations \eqref{eq:QLECase3FW} becomes
\begin{equation}
\begin{aligned}
\frac{\partial}{\partial z}\left[\begin{matrix}\hat a_{1}\\\hat a_2^\dag\\\end{matrix}\right]
=\left[\begin{matrix}0&i\kappa\\-i\kappa&0\\\end{matrix}\right]\left[\begin{matrix}\hat a_{1}\\{\hat a_2^{\dag}}\\\end{matrix}\right]
+\sqrt{\zeta}\left[\begin{matrix}1&1\\-1&1\\\end{matrix}\right]\left[\begin{matrix}\hat f_{1}\\{\hat f_2^{\dag}}\\\end{matrix}\right].
\end{aligned}\label{eq:QLECase3FWRabi}
\end{equation}
Under the condition $|\mathrm{M_F}L|\ll 1$, we solve Eq. ~\eqref{eq:QLECase3FWRabi} to the first order of $L$ and have
\begin{equation}
    \begin{aligned}
    &\hat a_1(L)\cong\hat a_1(0)+i\kappa L \hat a_2^{\dag}(0)+\sqrt{\zeta}\int_0^L dz \left(\hat f_1+\hat f_2^{\dag}\right),\\
    &\hat a_2(L)\cong\hat a_2(0)+i\kappa^* L \hat a_1^{\dag}(0)+\sqrt{\zeta}\int_0^L dz \left(-\hat f_1^{\dag}+\hat f_2\right).
    \end{aligned}
\end{equation}
The two-photon field correlations are
\begin{equation}
    \begin{aligned}
    \langle \hat a_1(L) \hat a_2(L)\rangle= \langle \hat a_2(L) \hat a_1(L)\rangle\cong \frac{i}{2}(\kappa+\kappa^*)L\delta(\varpi-\varpi').
    \end{aligned}
    \label{eq:positive}
\end{equation}
As $\zeta <0$, the Langevin coupled equations \eqref{eq:QLECase3FW} becomes
\begin{equation}
\begin{aligned}
\frac{\partial}{\partial z}\left[\begin{matrix}\hat a_{1}\\\hat a_2^\dag\\\end{matrix}\right]
=\left[\begin{matrix}0&i\kappa\\-i\kappa&0\\\end{matrix}\right]\left[\begin{matrix}\hat a_{1}\\{\hat a_2^{\dag}}\\\end{matrix}\right]
+\sqrt{-\zeta}\left[\begin{matrix}1&1\\-1&1\\\end{matrix}\right]\left[\begin{matrix}\hat f_{1}^{\dag}\\{\hat f_2}\\\end{matrix}\right].
\end{aligned}\label{eq:QLECase3FWRabi2}
\end{equation}
Under the condition $|\mathrm{M_F}L|\ll 1$, we solve Eq. ~\eqref{eq:QLECase3FWRabi2} to the first order of $L$ and have
\begin{equation}
    \begin{aligned}
    &\hat a_1(L)\cong\hat a_1(0)+i\kappa L \hat a_2^{\dag}(0)+\sqrt{-\zeta}\int_0^L dz \left(\hat f_1^{\dag}+\hat f_2\right),\\
    &\hat a_2(L)\cong\hat a_2(0)+i\kappa^* L \hat a_1^{\dag}(0)+\sqrt{-\zeta}\int_0^L dz \left(-\hat f_1+\hat f_2^{\dag}\right).
    \end{aligned}
\end{equation}
The two-photon field correlations are
\begin{equation}
    \begin{aligned}
    \langle \hat a_1(L) \hat a_2(L)\rangle=\langle \hat a_2(L) \hat a_1(L)\rangle\cong \frac{i}{2}(k+k^*)L\delta(\varpi-\varpi'),
    \end{aligned}
    \label{eq:negative}
\end{equation}
which is the same as Eq. \eqref{eq:positive}. The biphoton relative wavefunction is
\begin{equation}
\begin{aligned}
\psi_{21}(\tau)=\psi_{21}^*(-\tau)=\frac{iL}{2\pi}\int d\varpi \frac{1}{2}(k+k^*) e^{-i\varpi\tau}.
\end{aligned}
\label{eq:BiphtonRelativeWavefuncitonRabiFW02}
\end{equation}
One can prove that under the same limit $|\mathrm{M_B}L|\ll 1$, the backward-wave configuration gives the same two-photon field correlation [Eqs. \eqref{eq:positive} and \eqref{eq:negative}] and temporal wavefunction [Eq. \eqref{eq:BiphtonRelativeWavefuncitonRabiFW02}]. Equation \eqref{eq:BiphtonRelativeWavefuncitonRabiFW02} suggests the biphoton temporal wavefunction has time reversal symmetry when there is no linear gain and loss. 

\section{Conclusion} \label{sec:Conclusion}

In summary, we provide a macroscopic phenomenological formula of quantum Langevin equations for two coupled phase-conjugated fields with linear loss (gain) and complex nonlinear coupling coefficient, in both forward- and backward-wave configurations. The macroscopic phenomenological formula, obtained from the coupling matrix and the requirement of preserving commutation relations of field operators during propagation, does not require knowing microscopic details of light-matter interaction and internal atomic structures. To validate this phenomenological formula, we take SFWM in a double-$\Lambda$ four-level atomic system as an example to numerically confirm that our macroscopic phenomenological result is consistent with that obtained from microscopic Heisenberg-Langevin theory. As compared to the complicated microscopic theory which varies from system to system, the macroscopic coupled equations are much more friendly to experimentalists. We apply the quantum Langevin equations to study the effects of gain and/or loss as well as complex nonlinear coupling coefficient in biphoton generation, particularly to the temporal quantum correlations. We show that the computation complexity can be dramatically reduced by taking a proper order of field operators based on loss and gain. Making a comparison between the quantum Langevin theory (in the Heisenberg picture) and the perturbation theory (in the interaction picture \cite{du2008narrowband}), we extend the expression of complex phase mismatching to account for loss and gain. At last, we reveal Rabi oscillation in SFWM biphoton temporal correlation when the propagation effect is small. Although in this article we focus on biphoton generation from the spontaneous parametric process, the quantum Langevin coupled equations can also be used to study two-mode squeezing, parametric oscillation, and other quantum light state generation.

\begin{acknowledgments}
S.D. acknowledges support from DOE (DE-SC0022069), AFOSR (FA9550-22-1-0043) and NSF (CNS-2114076, 2228725).
\end{acknowledgments}

\appendix

\section{Noise Matrix in Backward-Wave Configuration} \label{Appendix: BW Noise Matrix}

In the macroscopic quantum Langevin equations, the requirement of preserving commutation relations allows multiple choices of the noise matrix. For example, $\hat f_1\rightarrow -\hat f_1$ or/and $\hat f_2\rightarrow -\hat f_2$ do not affect any computation results of physical observables involving pairs of Langevin noise operators. As an example, here we provide several equivalent noise matrices for backward-wave configuration:
\begin{equation}
\begin{aligned}
\mathrm{N_{B1}}&\equiv\left[\begin{matrix}1&0\\0&-1\\\end{matrix}\right]\sqrt{\left[\begin{matrix}-\mathrm{M_{B11}}&-\mathrm{M_{B12}}\\\mathrm{M_{B21}}&\mathrm{M_{B22}}\\\end{matrix}\right]+\left[\begin{matrix}-\mathrm{M_{B11}}&-\mathrm{M_{B12}}\\\mathrm{M_{B21}}&\mathrm{M_{B22}}\\\end{matrix}\right]^*}\\
&=\left[\begin{matrix}1&0\\0&-1\\\end{matrix}\right] \mathrm{N_{F}},\\
\mathrm{N_{B2}}&\equiv\mathrm{N_{B1}}\left[\begin{matrix}1&0\\0&-1\\\end{matrix}\right]\\
&=\sqrt{\left[\begin{matrix}-\mathrm{M_{B11}}&\mathrm{M_{B12}}\\-\mathrm{M_{B21}}&\mathrm{M_{B22}}\\\end{matrix}\right]+\left[\begin{matrix}-\mathrm{M_{B11}}&\mathrm{M_{B12}}\\-\mathrm{M_{B21}}&\mathrm{M_{B22}}\\\end{matrix}\right]^*},\\
\mathrm{N_{B3}}&\equiv\mathrm{N_{B1}}\left[\begin{matrix}-1&0\\0&1\\\end{matrix}\right],\\
\mathrm{N_{B4}}&\equiv\mathrm{N_{B1}}\left[\begin{matrix}-1&0\\0&-1\\\end{matrix}\right]=-\mathrm{N_{B1}}.
\end{aligned}\label{eq:NB1}
\end{equation}
We take the first choice $\mathrm{N_{B1}}$ in the main text [see Eq.~\eqref{eq:NoiseMatrixBackward} in Sec.~\ref{BW system}] so that it is consistent with the microscopic treatment in Sec.~\ref{sec:Microscopic}.

\section{Heisenberg-Langevin Equations of SFWM}\label{Appendix: Heisenberg-Langevin Equations in FWM}

The full Heisenberg equation of motion can be written as
\begin{equation}
\begin{aligned}
\dot {\hat{\mathcal{S}}}=i(\hat{\mathcal{O}}\hat{\mathcal{S}}-\hat{\mathcal{S}}\hat{\mathcal{O}})+\hat{\mathcal{G}}+\hat{\mathcal{F}},
\end{aligned}\label{eq:EoM mat}
\end{equation}
where
\begin{equation}
\begin{aligned}
\hat{\mathcal{S}}=\left[\begin{matrix}{\hat{\sigma}}_{11}&{\hat{\sigma}}_{12}&{\hat{\sigma}}_{13}&{\hat{\sigma}}_{14}
\\{\hat{\sigma}}_{21}&{\hat{\sigma}}_{22}&{\hat{\sigma}}_{23}&{\hat{\sigma}}_{24}
\\{\hat{\sigma}}_{31}&{\hat{\sigma}}_{32}&{\hat{\sigma}}_{33}&{\hat{\sigma}}_{34}
\\{\hat{\sigma}}_{41}&{\hat{\sigma}}_{42}&{\hat{\sigma}}_{43}&{\hat{\sigma}}_{44}\end{matrix}\right],
\end{aligned}\label{eq:Sigma mat}
\end{equation}
\begin{equation}
\begin{aligned}
\hat{\mathcal{O}}=-\left[\begin{matrix}0&0&g_{31}\hat{a}_{as}&\Omega_p/2
\\0&\varpi&\Omega_c/2&g_{42}\hat{a}_{s}
\\g_{13}\hat{a}_{as}^*&\Omega_c^*/2&\varpi&0
\\\Omega_p^*/2&g_{24}\hat{a}_{s}^*&0&\Delta_p\end{matrix}\right],
\end{aligned}\label{eq:H mat}
\end{equation}
\begin{equation}
\begin{aligned}
\hat{\mathcal{G}}=\left[\begin{matrix}\Gamma_{31}{\hat{\sigma}}_{33}+\Gamma_{41}{\hat{\sigma}}_{44}&-\gamma_{12}{\hat{\sigma}}_{12}&-\gamma_{13}{\hat{\sigma}}_{13}&-\gamma_{14}{\hat{\sigma}}_{14}
\\-\gamma_{12}{\hat{\sigma}}_{21}&\Gamma_{32}{\hat{\sigma}}_{33}+\Gamma_{42}{\hat{\sigma}}_{44}&-\gamma_{23}{\hat{\sigma}}_{23}&-\gamma_{24}{\hat{\sigma}}_{24}
\\-\gamma_{13}{\hat{\sigma}}_{31}&-\gamma_{23}{\hat{\sigma}}_{32}&-\Gamma_{3}{\hat{\sigma}}_{33}&-\gamma_{34}{\hat{\sigma}}_{34}
\\-\gamma_{14}{\hat{\sigma}}_{41}&-\gamma_{24}{\hat{\sigma}}_{42}&-\gamma_{34}{\hat{\sigma}}_{43}&-\Gamma_{4}{\hat{\sigma}}_{44}\end{matrix}\right],
\end{aligned}\label{eq:Gamma mat}
\end{equation}
\begin{equation}
\begin{aligned}
\hat{\mathcal{F}}=\left[\begin{matrix}{\hat{f}}_{11}^{(\sigma)}&{\hat{f}}_{12}^{(\sigma)}&{\hat{f}}_{13}^{(\sigma)}&{\hat{f}}_{14}^{(\sigma)}
\\{\hat{f}}_{21}^{(\sigma)}&{\hat{f}}_{22}^{(\sigma)}&{\hat{f}}_{23}^{(\sigma)}&{\hat{f}}_{24}^{(\sigma)}
\\{\hat{f}}_{31}^{(\sigma)}&{\hat{f}}_{32}^{(\sigma)}&{\hat{f}}_{33}^{(\sigma)}&{\hat{f}}_{34}^{(\sigma)}
\\{\hat{f}}_{41}^{(\sigma)}&{\hat{f}}_{42}^{(\sigma)}&{\hat{f}}_{43}^{(\sigma)}&{\hat{f}}_{44}^{(\sigma)}\end{matrix}\right].
\end{aligned}\label{eq:F mat}
\end{equation}
$\Gamma_m=\Gamma_{m1}+\Gamma_{m2}$ is the total spontaneous decay rate of excited state $\left|m\right\rangle$, where $m$ = 3, or 4, and $\Gamma_{mj}$ is the decay rate from state $\left|m\right\rangle$ to $\left|j\right\rangle$. For the two hyperfine ground states, there are $\Gamma_1=\Gamma_2=0$. For cold atoms with only spontaneous emisson decay, the dephasing rates $\gamma_{jk}$ ($j\neq k$) between states $\left|k\right\rangle$ and $\left|j\right\rangle$ are $\gamma_{13}=\gamma_{23}=\Gamma_3/2$, $\gamma_{14}=\gamma_{24}=\Gamma_4/2$, $\gamma_{34}=(\Gamma_3+\Gamma_4)/2$. $\gamma_{12}$ is the dephasing rate between two hyperfine ground states $\left|1\right\rangle$ and $\left|2\right\rangle$.

\section{Microscopic SFWM Quantum Langevin Equations in Forward-Wave Configuration} \label{Appendix: FW Quantum Langevin Equations}

Although Sec. \ref{sec:Microscopic} focuses on numerical confirmation of backward-wave SFWM, we remark that it may be helpful for general readers to write the SFWM quantum Langevin equations in the forward-wave configuration as well.

In the forward-wave configuration with both Stokes and anti-Stokes fields propagating along $+z$ direction, the coupled Langevin equations become
\begin{equation}
\begin{aligned}
\frac{\partial}{\partial z}\left[\begin{matrix}\hat a_{as}\\\hat a_s^\dag\\\end{matrix}\right]
=\mathrm{M_F}\left[\begin{matrix}\hat a_{as}\\{\hat a_s^{\dag}}\\\end{matrix}\right]+\left[\begin{matrix}\hat{F}_{as}\\ \hat {F}_s^{\dag}\\\end{matrix}\right],
\end{aligned}\label{eq:SFWM-FW}
\end{equation}
where
\begin{equation}
\begin{aligned}
\mathrm{M_F}=\left[\begin{matrix}-\alpha_{as}+i\frac{\Delta k}{2}&i\kappa\\-i\kappa&-\alpha_s^*-i\frac{\Delta k}{2}\\\end{matrix}\right],
\end{aligned}\label{eq:M-FW-SFWM}
\end{equation}
with $\Delta k=(\omega_{as}+\omega_{s})/c-(\vec{k}_c+\vec{k}_p)\cdot \hat{z}$. The noise operators $\hat F_{as}$ and $\hat F_s^\dag$, defined in Eq. \eqref{eq:F}, originate from microscopic atom-light interaction. To compare Eq. \eqref{eq:SFWM-FW} with Eq. \eqref{eq:QLEFW} from the phenomenological approach in Sec. \ref{sec:QLE}, we take mode 1 as anti-Stokes and mode 2 as Stokes in the forward-wave configuration. From Eq. \eqref{eq:QLEFW}, we can also obtain $\hat F_{as}$ and $\hat F_s^\dag$ from the noise matrix:
\begin{equation}
\begin{aligned}
&\hat F_{as}=
\mathrm{N_{FR}}_{11}\hat{f}_{1}
+\mathrm{N_{FI}}_{11}\hat{f}^{\dag}_{1}
+\mathrm{N_{FI}}_{12}\hat{f}_{2}
+\mathrm{N_{FR}}_{12}\hat{f}^{\dag}_{2},\\
&\hat F_{s}^\dag=
\mathrm{N_{FR}}_{21}\hat{f}_{1}
+\mathrm{N_{FI}}_{21}\hat{f}^{\dag}_{1}
+\mathrm{N_{FI}}_{22}\hat{f}_{2}
+\mathrm{N_{FR}}_{22}\hat{f}^{\dag}_{2}.
\end{aligned}\label{eq:FpFW}
\end{equation}

\section{Complex Phase Mismatching in Forward-Wave Configuration} \label{Appendix: FW Complex Phase Mismatching}

In the forward-wave configuration, similar to the backward-wave configuration in Sec.~\ref{Complex Phase Mismatching}, we assume anti-Stokes photons in mode 1 are lossless with EIT and there is gain (or loss) in Stokes mode 2. The small parametric gain fulfills $\left|\kappa\right| \ll \{\alpha, g\}$. Using Eq. (\ref{eq:MFW}) and (\ref{ABCDFW}), we obtain analytical expressions of $A, B, C$, and $D$ as
\begin{equation}
\begin{aligned}
A&= \frac{\sqrt{q^2+4\kappa^2}\mathrm{cosh}\left(\frac{L}{2}\sqrt{q^2+4\kappa^2}\right)-q\mathrm{sinh}\left(\frac{L}{2}\sqrt{q^2+4\kappa^2}\right)}{\sqrt{q^2+4\kappa^2}e^{(\alpha_1+\alpha_2^*)L/2}},\\
B&= \frac{2i\kappa\mathrm{sinh}\left(\frac{L}{2}\sqrt{q^2+4\kappa^2}\right)}{\sqrt{q^2+4\kappa^2}e^{(\alpha_1+\alpha_2^*)L/2}},\\
C&= \frac{-2i\kappa\mathrm{sinh}\left(\frac{L}{2}\sqrt{q^2+4\kappa^2}\right)}{\sqrt{q^2+4\kappa^2}e^{(\alpha_1+\alpha_2^*)L/2}},\\
D&= \frac{\sqrt{q^2+4\kappa^2}\mathrm{cosh}\left(\frac{L}{2}\sqrt{q^2+4\kappa^2}\right)+q\mathrm{sinh}\left(\frac{L}{2}\sqrt{q^2+4\kappa^2}\right)}{\sqrt{q^2+4\kappa^2}e^{(\alpha_1+\alpha_2^*)L/2}},
\end{aligned}\label{ABCDexprFW}
\end{equation}
where $q\equiv\alpha_1-\alpha_2^*-i\Delta k$. In the small parametric gain approximation, we have
\begin{equation}
\begin{aligned}
&\sqrt{q^2-4\kappa^2}\approx q\\ &=\alpha_1-\alpha_2^*-i\Delta k
=-i(\Delta k_1+\Delta k_2^*+\Delta k),
\end{aligned}
\end{equation}
and
\begin{equation}
\begin{aligned}
\alpha_1+\alpha_2^*
=-i(\Delta k_1-\Delta k_2^*),
\end{aligned}
\end{equation} 
where $\Delta k_m=\frac{\omega_m}{2c}\chi_{m}$ is the wavenumber difference from that in vacuum. Hence, we simplify $A, B, C$, and $D$ to
\begin{equation}
\begin{aligned}
A=&\mathrm{exp}\left[i\Delta k_1 L\right]\mathrm{exp}\left[\frac{i\Delta k L}{2}\right],\\
B=&i\kappa L \mathrm{sinc}\left[\frac{(\Delta k_1+\Delta k_2^*+\Delta k)L}{2}\right]\\
&\times\mathrm{exp}\left[\frac{i(\Delta k_1-\Delta k_2^*)L}{2}\right],\\
C=&-i\kappa L \mathrm{sinc}\left[\frac{(\Delta k_1+\Delta k_2^*+\Delta k)L}{2}\right]\\
&\times\mathrm{exp}\left[\frac{i(\Delta k_1-\Delta k_2^*)L}{2}\right],\\
D=&\mathrm{exp}\left[-i\Delta k_2^* L\right]\mathrm{exp}\left[\frac{-i\Delta k L}{2}\right].
\end{aligned}\label{ABCDexpr2FW}
\end{equation}

We first look at the case with gain in the Stokes (mode 2). As discussed in Sec. \ref{sec:TimeOrder}, we take the order $\langle:\hat a_{2}\hat a_{1}:\rangle$ 
\begin{equation}
\begin{aligned}
\psi_{21}(\tau)&=\iint d\varpi d\varpi' \langle \hat a_{2,out}(\varpi')\hat a_{1,out}(\varpi)\rangle e^{-i\varpi \tau}\\
&=\int d\varpi BD^*e^{-i\varpi \tau},
\end{aligned}
\label{eq:psigain1FW}
\end{equation}
where
\begin{equation}
\begin{aligned}
{{B}}{{D}^\ast}
&=i\kappa L\mathrm{sinc}\left[\frac{(\Delta k_1+\Delta k_2^*+\Delta k)L}{2}\right]\\
&\times\mathrm{exp}\left[\frac{i(\Delta k_1-\Delta k_2^*+2\Delta k_2+\Delta k)L}{2}\right].
\end{aligned}\label{eq:BDgainFW}
\end{equation}
Comparing Eqs.~\eqref{eq:psigain1FW} and \eqref{eq:BDgainFW} with Eqs.~\eqref{eq:psiJOSAB} and \eqref{eq:PhiJOSAB}, particularly for the argument in the $\mathrm{sinc}$ function, we have $\Delta\tilde k=\Delta k_1+\Delta k_2^*+\Delta k=k_{1}+k_2^*-k_c-k_p=k_{as}+k_s^*-k_c-k_p$ which is consistent with the statement in Ref.~\cite{du2008narrowband}. 

We now look at the case with loss in the Stokes (mode 2). We take the order $\langle:\hat a_{1}\hat a_{2}:\rangle$ and have 
\begin{equation}
\begin{aligned}
\psi_{12}(\tau)&=\iint d\varpi d\varpi' \langle \hat a_{1,out}(\varpi)\hat a_{2,out}(\varpi')\rangle e^{-i\varpi \tau}\\
&=\int d\varpi AC^*e^{-i\varpi \tau},
\end{aligned}
\label{eq:psiloss1FW}
\end{equation}
where
\begin{equation}
\begin{aligned}
{{A}}{{C}^\ast}&=i\kappa^\ast L\mathrm{sinc}\left[\frac{(\Delta k_1^*+\Delta k_2+\Delta k)L}{2}\right]\\
&\times\mathrm{exp}\left[\frac{i(2\Delta k_1-\Delta k_1^*+\Delta k_2+\Delta k)L}{2}\right].
\end{aligned}\label{eq:AClossFW}
\end{equation}
Comparing Eqs.~\eqref{eq:psiloss1FW} and \eqref{eq:AClossFW} with Eqs.~\eqref{eq:psiJOSAB} and \eqref{eq:PhiJOSAB}, we have $\Delta\tilde k=\Delta k_1^*+\Delta k_2+\Delta k=k_{1}+k_2-k_c+k_p=k_{as}+k_s-k_c-k_p$, which is different from the case with gain. Here we have taken $k_1\simeq k_1^*$ for lossless mode 1. Therefore, in the case with loss in the Stokes mode 2, the complex phase mismatching becomes $\Delta\tilde k = \left(\vec{k}_{as}+\vec{k}_s-\vec{k}_c-\vec{k}_p \right)\cdot \hat{z}$.

\bibliography{Langevin}

\begin{thebibliography}{25}%
\makeatletter
\providecommand \@ifxundefined [1]{%
 \@ifx{#1\undefined}
}%
\providecommand \@ifnum [1]{%
 \ifnum #1\expandafter \@firstoftwo
 \else \expandafter \@secondoftwo
 \fi
}%
\providecommand \@ifx [1]{%
 \ifx #1\expandafter \@firstoftwo
 \else \expandafter \@secondoftwo
 \fi
}%
\providecommand \natexlab [1]{#1}%
\providecommand \enquote  [1]{``#1''}%
\providecommand \bibnamefont  [1]{#1}%
\providecommand \bibfnamefont [1]{#1}%
\providecommand \citenamefont [1]{#1}%
\providecommand \href@noop [0]{\@secondoftwo}%
\providecommand \href [0]{\begingroup \@sanitize@url \@href}%
\providecommand \@href[1]{\@@startlink{#1}\@@href}%
\providecommand \@@href[1]{\endgroup#1\@@endlink}%
\providecommand \@sanitize@url [0]{\catcode `\\12\catcode `\$12\catcode
  `\&12\catcode `\#12\catcode `\^12\catcode `\_12\catcode `\%12\relax}%
\providecommand \@@startlink[1]{}%
\providecommand \@@endlink[0]{}%
\providecommand \url  [0]{\begingroup\@sanitize@url \@url }%
\providecommand \@url [1]{\endgroup\@href {#1}{\urlprefix }}%
\providecommand \urlprefix  [0]{URL }%
\providecommand \Eprint [0]{\href }%
\providecommand \doibase [0]{https://doi.org/}%
\providecommand \selectlanguage [0]{\@gobble}%
\providecommand \bibinfo  [0]{\@secondoftwo}%
\providecommand \bibfield  [0]{\@secondoftwo}%
\providecommand \translation [1]{[#1]}%
\providecommand \BibitemOpen [0]{}%
\providecommand \bibitemStop [0]{}%
\providecommand \bibitemNoStop [0]{.\EOS\space}%
\providecommand \EOS [0]{\spacefactor3000\relax}%
\providecommand \BibitemShut  [1]{\csname bibitem#1\endcsname}%
\let\auto@bib@innerbib\@empty
\bibitem [{\citenamefont {Gardiner}\ and\ \citenamefont
  {Collett}(1985)}]{gardiner1985input}%
  \BibitemOpen
  \bibfield  {author} {\bibinfo {author} {\bibfnamefont {C.~W.}\ \bibnamefont
  {Gardiner}}\ and\ \bibinfo {author} {\bibfnamefont {M.~J.}\ \bibnamefont
  {Collett}},\ }\bibfield  {title} {\bibinfo {title} {Input and output in
  damped quantum systems: Quantum stochastic differential equations and the
  master equation},\ }\href {https://doi.org/10.1103/PhysRevA.31.3761}
  {\bibfield  {journal} {\bibinfo  {journal} {Phys. Rev. A}\ }\textbf {\bibinfo
  {volume} {31}},\ \bibinfo {pages} {3761} (\bibinfo {year}
  {1985})}\BibitemShut {NoStop}%
\bibitem [{\citenamefont {Scully}\ and\ \citenamefont
  {Zubairy}(1999)}]{scully1999quantum}%
  \BibitemOpen
  \bibfield  {author} {\bibinfo {author} {\bibfnamefont {M.~O.}\ \bibnamefont
  {Scully}}\ and\ \bibinfo {author} {\bibfnamefont {M.~S.}\ \bibnamefont
  {Zubairy}},\ }\href@noop {} {\bibinfo {title} {Quantum optics}} (\bibinfo
  {year} {1999})\BibitemShut {NoStop}%
\bibitem [{\citenamefont {Yamamoto}\ and\ \citenamefont
  {Imamoglu}(1999)}]{yamamoto1999mesoscopic}%
  \BibitemOpen
  \bibfield  {author} {\bibinfo {author} {\bibfnamefont {Y.}~\bibnamefont
  {Yamamoto}}\ and\ \bibinfo {author} {\bibfnamefont {A.}~\bibnamefont
  {Imamoglu}},\ }\bibfield  {title} {\bibinfo {title} {Mesoscopic quantum
  optics},\ }\href@noop {} {\bibfield  {journal} {\bibinfo  {journal}
  {Mesoscopic Quantum Optics}\ } (\bibinfo {year} {1999})}\BibitemShut
  {NoStop}%
\bibitem [{\citenamefont {Gardiner}\ \emph {et~al.}(2004)\citenamefont
  {Gardiner}, \citenamefont {Zoller},\ and\ \citenamefont
  {Zoller}}]{gardiner2004quantum}%
  \BibitemOpen
  \bibfield  {author} {\bibinfo {author} {\bibfnamefont {C.}~\bibnamefont
  {Gardiner}}, \bibinfo {author} {\bibfnamefont {P.}~\bibnamefont {Zoller}},\
  and\ \bibinfo {author} {\bibfnamefont {P.}~\bibnamefont {Zoller}},\
  }\href@noop {} {\emph {\bibinfo {title} {Quantum noise: a handbook of
  Markovian and non-Markovian quantum stochastic methods with applications to
  quantum optics}}}\ (\bibinfo  {publisher} {Springer Science \& Business
  Media},\ \bibinfo {year} {2004})\BibitemShut {NoStop}%
\bibitem [{\citenamefont {Benguria}\ and\ \citenamefont
  {Kac}(1981)}]{PhysRevLett.46.1}%
  \BibitemOpen
  \bibfield  {author} {\bibinfo {author} {\bibfnamefont {R.}~\bibnamefont
  {Benguria}}\ and\ \bibinfo {author} {\bibfnamefont {M.}~\bibnamefont {Kac}},\
  }\bibfield  {title} {\bibinfo {title} {Quantum langevin equation},\ }\href
  {https://doi.org/10.1103/PhysRevLett.46.1} {\bibfield  {journal} {\bibinfo
  {journal} {Phys. Rev. Lett.}\ }\textbf {\bibinfo {volume} {46}},\ \bibinfo
  {pages} {1} (\bibinfo {year} {1981})}\BibitemShut {NoStop}%
\bibitem [{\citenamefont {Boyd}(2020)}]{boyd2020nonlinear}%
  \BibitemOpen
  \bibfield  {author} {\bibinfo {author} {\bibfnamefont {R.~W.}\ \bibnamefont
  {Boyd}},\ }\href@noop {} {\emph {\bibinfo {title} {Nonlinear optics}}}\
  (\bibinfo  {publisher} {Academic press},\ \bibinfo {year} {2020})\BibitemShut
  {NoStop}%
\bibitem [{\citenamefont {Shwartz}\ \emph {et~al.}(2012)\citenamefont
  {Shwartz}, \citenamefont {Coffee}, \citenamefont {Feldkamp}, \citenamefont
  {Feng}, \citenamefont {Hastings}, \citenamefont {Yin},\ and\ \citenamefont
  {Harris}}]{shwartz2012x}%
  \BibitemOpen
  \bibfield  {author} {\bibinfo {author} {\bibfnamefont {S.}~\bibnamefont
  {Shwartz}}, \bibinfo {author} {\bibfnamefont {R.~N.}\ \bibnamefont {Coffee}},
  \bibinfo {author} {\bibfnamefont {J.~M.}\ \bibnamefont {Feldkamp}}, \bibinfo
  {author} {\bibfnamefont {Y.}~\bibnamefont {Feng}}, \bibinfo {author}
  {\bibfnamefont {J.~B.}\ \bibnamefont {Hastings}}, \bibinfo {author}
  {\bibfnamefont {G.~Y.}\ \bibnamefont {Yin}},\ and\ \bibinfo {author}
  {\bibfnamefont {S.~E.}\ \bibnamefont {Harris}},\ }\bibfield  {title}
  {\bibinfo {title} {X-ray parametric down-conversion in the langevin regime},\
  }\href {https://doi.org/10.1103/PhysRevLett.109.013602} {\bibfield  {journal}
  {\bibinfo  {journal} {Phys. Rev. Lett.}\ }\textbf {\bibinfo {volume} {109}},\
  \bibinfo {pages} {013602} (\bibinfo {year} {2012})}\BibitemShut {NoStop}%
\bibitem [{\citenamefont {Javid}\ and\ \citenamefont
  {Lin}(2019)}]{PhysRevA.100.043811}%
  \BibitemOpen
  \bibfield  {author} {\bibinfo {author} {\bibfnamefont {U.~A.}\ \bibnamefont
  {Javid}}\ and\ \bibinfo {author} {\bibfnamefont {Q.}~\bibnamefont {Lin}},\
  }\bibfield  {title} {\bibinfo {title} {Quantum correlations from dynamically
  modulated optical nonlinear interactions},\ }\href
  {https://doi.org/10.1103/PhysRevA.100.043811} {\bibfield  {journal} {\bibinfo
   {journal} {Phys. Rev. A}\ }\textbf {\bibinfo {volume} {100}},\ \bibinfo
  {pages} {043811} (\bibinfo {year} {2019})}\BibitemShut {NoStop}%
\bibitem [{\citenamefont {Shafiee}\ \emph {et~al.}(2020)\citenamefont
  {Shafiee}, \citenamefont {Strekalov}, \citenamefont {Otterpohl},
  \citenamefont {Sedlmeir}, \citenamefont {Schunk}, \citenamefont {Vogl},
  \citenamefont {Schwefel}, \citenamefont {Leuchs},\ and\ \citenamefont
  {Marquardt}}]{Shafiee_2020}%
  \BibitemOpen
  \bibfield  {author} {\bibinfo {author} {\bibfnamefont {G.}~\bibnamefont
  {Shafiee}}, \bibinfo {author} {\bibfnamefont {D.~V.}\ \bibnamefont
  {Strekalov}}, \bibinfo {author} {\bibfnamefont {A.}~\bibnamefont
  {Otterpohl}}, \bibinfo {author} {\bibfnamefont {F.}~\bibnamefont {Sedlmeir}},
  \bibinfo {author} {\bibfnamefont {G.}~\bibnamefont {Schunk}}, \bibinfo
  {author} {\bibfnamefont {U.}~\bibnamefont {Vogl}}, \bibinfo {author}
  {\bibfnamefont {H.~G.~L.}\ \bibnamefont {Schwefel}}, \bibinfo {author}
  {\bibfnamefont {G.}~\bibnamefont {Leuchs}},\ and\ \bibinfo {author}
  {\bibfnamefont {C.}~\bibnamefont {Marquardt}},\ }\bibfield  {title} {\bibinfo
  {title} {Nonlinear power dependence of the spectral properties of an optical
  parametric oscillator below threshold in the quantum regime},\ }\href
  {https://doi.org/10.1088/1367-2630/ab9a87} {\bibfield  {journal} {\bibinfo
  {journal} {New Journal of Physics}\ }\textbf {\bibinfo {volume} {22}},\
  \bibinfo {pages} {073045} (\bibinfo {year} {2020})}\BibitemShut {NoStop}%
\bibitem [{\citenamefont {Du}\ \emph {et~al.}(2008)\citenamefont {Du},
  \citenamefont {Wen},\ and\ \citenamefont {Rubin}}]{du2008narrowband}%
  \BibitemOpen
  \bibfield  {author} {\bibinfo {author} {\bibfnamefont {S.}~\bibnamefont
  {Du}}, \bibinfo {author} {\bibfnamefont {J.}~\bibnamefont {Wen}},\ and\
  \bibinfo {author} {\bibfnamefont {M.~H.}\ \bibnamefont {Rubin}},\ }\bibfield
  {title} {\bibinfo {title} {Narrowband biphoton generation near atomic
  resonance},\ }\href {https://doi.org/10.1364/JOSAB.25.000C98} {\bibfield
  {journal} {\bibinfo  {journal} {J. Opt. Soc. Am. B}\ }\textbf {\bibinfo
  {volume} {25}},\ \bibinfo {pages} {C98} (\bibinfo {year} {2008})}\BibitemShut
  {NoStop}%
\bibitem [{\citenamefont {Kolchin}(2007)}]{kolchin2007electromagnetically}%
  \BibitemOpen
  \bibfield  {author} {\bibinfo {author} {\bibfnamefont {P.}~\bibnamefont
  {Kolchin}},\ }\bibfield  {title} {\bibinfo {title}
  {Electromagnetically-induced-transparency-based paired photon generation},\
  }\href {https://doi.org/10.1103/PhysRevA.75.033814} {\bibfield  {journal}
  {\bibinfo  {journal} {Phys. Rev. A}\ }\textbf {\bibinfo {volume} {75}},\
  \bibinfo {pages} {033814} (\bibinfo {year} {2007})}\BibitemShut {NoStop}%
\bibitem [{\citenamefont {Zhao}\ \emph {et~al.}(2016)\citenamefont {Zhao},
  \citenamefont {Su},\ and\ \citenamefont {Du}}]{zhao2016narrowband}%
  \BibitemOpen
  \bibfield  {author} {\bibinfo {author} {\bibfnamefont {L.}~\bibnamefont
  {Zhao}}, \bibinfo {author} {\bibfnamefont {Y.}~\bibnamefont {Su}},\ and\
  \bibinfo {author} {\bibfnamefont {S.}~\bibnamefont {Du}},\ }\bibfield
  {title} {\bibinfo {title} {Narrowband biphoton generation in the group delay
  regime},\ }\href {https://doi.org/10.1103/PhysRevA.93.033815} {\bibfield
  {journal} {\bibinfo  {journal} {Phys. Rev. A}\ }\textbf {\bibinfo {volume}
  {93}},\ \bibinfo {pages} {033815} (\bibinfo {year} {2016})}\BibitemShut
  {NoStop}%
\bibitem [{\citenamefont {Raymond~Ooi}\ \emph {et~al.}(2007)\citenamefont
  {Raymond~Ooi}, \citenamefont {Sun}, \citenamefont {Zubairy},\ and\
  \citenamefont {Scully}}]{ooi2007correlation}%
  \BibitemOpen
  \bibfield  {author} {\bibinfo {author} {\bibfnamefont {C.~H.}\ \bibnamefont
  {Raymond~Ooi}}, \bibinfo {author} {\bibfnamefont {Q.}~\bibnamefont {Sun}},
  \bibinfo {author} {\bibfnamefont {M.~S.}\ \bibnamefont {Zubairy}},\ and\
  \bibinfo {author} {\bibfnamefont {M.~O.}\ \bibnamefont {Scully}},\ }\bibfield
   {title} {\bibinfo {title} {Correlation of photon pairs from the double raman
  amplifier: Generalized analytical quantum langevin theory},\ }\href
  {https://doi.org/10.1103/PhysRevA.75.013820} {\bibfield  {journal} {\bibinfo
  {journal} {Phys. Rev. A}\ }\textbf {\bibinfo {volume} {75}},\ \bibinfo
  {pages} {013820} (\bibinfo {year} {2007})}\BibitemShut {NoStop}%
\bibitem [{\citenamefont {Jiang}\ \emph {et~al.}(2019)\citenamefont {Jiang},
  \citenamefont {Mei}, \citenamefont {Zuo}, \citenamefont {Zhai}, \citenamefont
  {Li}, \citenamefont {Wen},\ and\ \citenamefont
  {Du}}]{PhysRevLett.123.193604}%
  \BibitemOpen
  \bibfield  {author} {\bibinfo {author} {\bibfnamefont {Y.}~\bibnamefont
  {Jiang}}, \bibinfo {author} {\bibfnamefont {Y.}~\bibnamefont {Mei}}, \bibinfo
  {author} {\bibfnamefont {Y.}~\bibnamefont {Zuo}}, \bibinfo {author}
  {\bibfnamefont {Y.}~\bibnamefont {Zhai}}, \bibinfo {author} {\bibfnamefont
  {J.}~\bibnamefont {Li}}, \bibinfo {author} {\bibfnamefont {J.}~\bibnamefont
  {Wen}},\ and\ \bibinfo {author} {\bibfnamefont {S.}~\bibnamefont {Du}},\
  }\bibfield  {title} {\bibinfo {title} {Anti-parity-time symmetric optical
  four-wave mixing in cold atoms},\ }\href
  {https://doi.org/10.1103/PhysRevLett.123.193604} {\bibfield  {journal}
  {\bibinfo  {journal} {Phys. Rev. Lett.}\ }\textbf {\bibinfo {volume} {123}},\
  \bibinfo {pages} {193604} (\bibinfo {year} {2019})}\BibitemShut {NoStop}%
\bibitem [{\citenamefont {Mei}\ \emph {et~al.}(2017)\citenamefont {Mei},
  \citenamefont {Guo}, \citenamefont {Zhao},\ and\ \citenamefont
  {Du}}]{mei2017mirrorless}%
  \BibitemOpen
  \bibfield  {author} {\bibinfo {author} {\bibfnamefont {Y.}~\bibnamefont
  {Mei}}, \bibinfo {author} {\bibfnamefont {X.}~\bibnamefont {Guo}}, \bibinfo
  {author} {\bibfnamefont {L.}~\bibnamefont {Zhao}},\ and\ \bibinfo {author}
  {\bibfnamefont {S.}~\bibnamefont {Du}},\ }\bibfield  {title} {\bibinfo
  {title} {Mirrorless optical parametric oscillation with tunable threshold in
  cold atoms},\ }\href {https://doi.org/10.1103/PhysRevLett.119.150406}
  {\bibfield  {journal} {\bibinfo  {journal} {Phys. Rev. Lett.}\ }\textbf
  {\bibinfo {volume} {119}},\ \bibinfo {pages} {150406} (\bibinfo {year}
  {2017})}\BibitemShut {NoStop}%
\bibitem [{\citenamefont {Luo}\ \emph {et~al.}(2022)\citenamefont {Luo},
  \citenamefont {Zhang},\ and\ \citenamefont {Du}}]{PhysRevLett.128.173602}%
  \BibitemOpen
  \bibfield  {author} {\bibinfo {author} {\bibfnamefont {X.-W.}\ \bibnamefont
  {Luo}}, \bibinfo {author} {\bibfnamefont {C.}~\bibnamefont {Zhang}},\ and\
  \bibinfo {author} {\bibfnamefont {S.}~\bibnamefont {Du}},\ }\bibfield
  {title} {\bibinfo {title} {Quantum squeezing and sensing with
  pseudo-anti-parity-time symmetry},\ }\href
  {https://doi.org/10.1103/PhysRevLett.128.173602} {\bibfield  {journal}
  {\bibinfo  {journal} {Phys. Rev. Lett.}\ }\textbf {\bibinfo {volume} {128}},\
  \bibinfo {pages} {173602} (\bibinfo {year} {2022})}\BibitemShut {NoStop}%
\bibitem [{\citenamefont {Bender}\ and\ \citenamefont
  {Boettcher}(1998)}]{PhysRevLett.80.5243}%
  \BibitemOpen
  \bibfield  {author} {\bibinfo {author} {\bibfnamefont {C.~M.}\ \bibnamefont
  {Bender}}\ and\ \bibinfo {author} {\bibfnamefont {S.}~\bibnamefont
  {Boettcher}},\ }\bibfield  {title} {\bibinfo {title} {Real spectra in
  non-{H}ermitian {H}amiltonians having {PT} symmetry},\ }\href
  {https://doi.org/10.1103/PhysRevLett.80.5243} {\bibfield  {journal} {\bibinfo
   {journal} {Phys. Rev. Lett.}\ }\textbf {\bibinfo {volume} {80}},\ \bibinfo
  {pages} {5243} (\bibinfo {year} {1998})}\BibitemShut {NoStop}%
\bibitem [{\citenamefont {Miri}\ and\ \citenamefont
  {Al{\`{u}}}(2019)}]{Miri_2019}%
  \BibitemOpen
  \bibfield  {author} {\bibinfo {author} {\bibfnamefont {M.-A.}\ \bibnamefont
  {Miri}}\ and\ \bibinfo {author} {\bibfnamefont {A.}~\bibnamefont
  {Al{\`{u}}}},\ }\bibfield  {title} {\bibinfo {title} {Exceptional points in
  optics and photonics},\ }\bibfield  {journal} {\bibinfo  {journal} {Science}\
  }\textbf {\bibinfo {volume} {363}},\ \href
  {https://doi.org/10.1126/science.aar7709} {10.1126/science.aar7709} (\bibinfo
  {year} {2019})\BibitemShut {NoStop}%
\bibitem [{\citenamefont {Braje}\ \emph {et~al.}(2004)\citenamefont {Braje},
  \citenamefont {Bali\ifmmode~\acute{c}\else \'{c}\fi{}}, \citenamefont {Goda},
  \citenamefont {Yin},\ and\ \citenamefont {Harris}}]{PhysRevLett.93.183601}%
  \BibitemOpen
  \bibfield  {author} {\bibinfo {author} {\bibfnamefont {D.~A.}\ \bibnamefont
  {Braje}}, \bibinfo {author} {\bibfnamefont {V.}~\bibnamefont
  {Bali\ifmmode~\acute{c}\else \'{c}\fi{}}}, \bibinfo {author} {\bibfnamefont
  {S.}~\bibnamefont {Goda}}, \bibinfo {author} {\bibfnamefont {G.~Y.}\
  \bibnamefont {Yin}},\ and\ \bibinfo {author} {\bibfnamefont {S.~E.}\
  \bibnamefont {Harris}},\ }\bibfield  {title} {\bibinfo {title} {Frequency
  mixing using electromagnetically induced transparency in cold atoms},\ }\href
  {https://doi.org/10.1103/PhysRevLett.93.183601} {\bibfield  {journal}
  {\bibinfo  {journal} {Phys. Rev. Lett.}\ }\textbf {\bibinfo {volume} {93}},\
  \bibinfo {pages} {183601} (\bibinfo {year} {2004})}\BibitemShut {NoStop}%
\bibitem [{\citenamefont {Bali\ifmmode~\acute{c}\else \'{c}\fi{}}\ \emph
  {et~al.}(2005)\citenamefont {Bali\ifmmode~\acute{c}\else \'{c}\fi{}},
  \citenamefont {Braje}, \citenamefont {Kolchin}, \citenamefont {Yin},\ and\
  \citenamefont {Harris}}]{balic2005generation}%
  \BibitemOpen
  \bibfield  {author} {\bibinfo {author} {\bibfnamefont {V.}~\bibnamefont
  {Bali\ifmmode~\acute{c}\else \'{c}\fi{}}}, \bibinfo {author} {\bibfnamefont
  {D.~A.}\ \bibnamefont {Braje}}, \bibinfo {author} {\bibfnamefont
  {P.}~\bibnamefont {Kolchin}}, \bibinfo {author} {\bibfnamefont {G.~Y.}\
  \bibnamefont {Yin}},\ and\ \bibinfo {author} {\bibfnamefont {S.~E.}\
  \bibnamefont {Harris}},\ }\bibfield  {title} {\bibinfo {title} {Generation of
  paired photons with controllable waveforms},\ }\href
  {https://doi.org/10.1103/PhysRevLett.94.183601} {\bibfield  {journal}
  {\bibinfo  {journal} {Phys. Rev. Lett.}\ }\textbf {\bibinfo {volume} {94}},\
  \bibinfo {pages} {183601} (\bibinfo {year} {2005})}\BibitemShut {NoStop}%
\bibitem [{\citenamefont {Harris}(1997)}]{EIT-Harris}%
  \BibitemOpen
  \bibfield  {author} {\bibinfo {author} {\bibfnamefont {S.~E.}\ \bibnamefont
  {Harris}},\ }\bibfield  {title} {\bibinfo {title} {Electromagnetically
  induced transparency},\ }\href@noop {} {\bibfield  {journal} {\bibinfo
  {journal} {Phys. Today}\ }\textbf {\bibinfo {volume} {50}},\ \bibinfo {pages}
  {36} (\bibinfo {year} {1997})}\BibitemShut {NoStop}%
\bibitem [{\citenamefont {Fleischhauer}\ \emph {et~al.}(2005)\citenamefont
  {Fleischhauer}, \citenamefont {Imamoglu},\ and\ \citenamefont
  {Marangos}}]{RevModPhys.77.633}%
  \BibitemOpen
  \bibfield  {author} {\bibinfo {author} {\bibfnamefont {M.}~\bibnamefont
  {Fleischhauer}}, \bibinfo {author} {\bibfnamefont {A.}~\bibnamefont
  {Imamoglu}},\ and\ \bibinfo {author} {\bibfnamefont {J.~P.}\ \bibnamefont
  {Marangos}},\ }\bibfield  {title} {\bibinfo {title} {Electromagnetically
  induced transparency: Optics in coherent media},\ }\href
  {https://doi.org/10.1103/RevModPhys.77.633} {\bibfield  {journal} {\bibinfo
  {journal} {Rev. Mod. Phys.}\ }\textbf {\bibinfo {volume} {77}},\ \bibinfo
  {pages} {633} (\bibinfo {year} {2005})}\BibitemShut {NoStop}%
\bibitem [{\citenamefont {Mandel}\ and\ \citenamefont {Wolf}(1994)}]{Mandel}%
  \BibitemOpen
  \bibfield  {author} {\bibinfo {author} {\bibfnamefont {L.}~\bibnamefont
  {Mandel}}\ and\ \bibinfo {author} {\bibfnamefont {E.}~\bibnamefont {Wolf}},\
  }\href@noop {} {\emph {\bibinfo {title} {Optical Coherence and Quantum
  Optics}}}\ (\bibinfo  {publisher} {Cambridge University Press, Cambridge
  England, New York},\ \bibinfo {year} {1994})\BibitemShut {NoStop}%
\bibitem [{\citenamefont {Louisell}(1973)}]{Louisell}%
  \BibitemOpen
  \bibfield  {author} {\bibinfo {author} {\bibfnamefont {W.~H.}\ \bibnamefont
  {Louisell}},\ }\href@noop {} {\emph {\bibinfo {title} {Optical Coherence and
  Quantum Optics}}}\ (\bibinfo  {publisher} {Wiley, New York},\ \bibinfo {year}
  {1973})\BibitemShut {NoStop}%
\bibitem [{\citenamefont {Du}(2015)}]{PhysRevA.92.043836}%
  \BibitemOpen
  \bibfield  {author} {\bibinfo {author} {\bibfnamefont {S.}~\bibnamefont
  {Du}},\ }\bibfield  {title} {\bibinfo {title} {Quantum-state purity of
  heralded single photons produced from frequency-anticorrelated biphotons},\
  }\href {https://doi.org/10.1103/PhysRevA.92.043836} {\bibfield  {journal}
  {\bibinfo  {journal} {Phys. Rev. A}\ }\textbf {\bibinfo {volume} {92}},\
  \bibinfo {pages} {043836} (\bibinfo {year} {2015})}\BibitemShut {NoStop}%
\end{thebibliography}%

\end{document}